Imperial College London

*BSc Research Project Report*

# A Computational Study and Parameterisation of Phosphatidylinositol Phosphates for Lipid Simulations with AMBER

*Author:*

Clare Xie Yijia

BSc Chemistry, Year 3

*Supervisor:*

Prof. Ian Gould

IMPERIAL COLLEGE LONDON

DEPARTMENT OF CHEMISTRY

June 2021

# Abstract

Phosphatidylinositol phosphates (PIPs) are membrane phospholipids that play crucial roles in a wide range of cellular functions. However, there is a dearth of experimental data in the literature on PIPs because their investigations are not always feasible and often prohibitively expensive. Hence, there is great interest in using computational simulations to study the structures, transitions, and interactions of PIP-containing membranes.

Assisted Model Building with Energy Refinement (AMBER) is a molecular dynamics program with validated force fields parameterised for a range of lipid types. The development of AMBER's Lipid force fields is a continuous process with ongoing work in both improving the accuracy of the simulated result and extending the force field with diverse lipid types. The research conducted for this work represents the first attempt to parameterise PIPs using the protocols of the latest Lipid21 force field.

PIPs' biological functions depend on the number, position, and ionisation state of the phosphate groups contained in their inositol rings. In this study, all 7 known biologically-active PIPs were studied with the consideration of every possible phosphate charge states, accounting for a total of 26 stereochemical structures. The geometry of each structure was optimised in Gaussian in 3 steps with MP2/6-31g*, MP2/cc-pVDZ, and finally MP2/cc-pVTZ. The electrostatic potential (ESP) was then calculated at the MP2/cc-pVTZ level and used to derive partial charges for each PIP structure using a two-stage restrained electrostatic potential (RESP) fit.

The relative energetic stability of the optimised structures were investigated and compared with available literature values, providing validation to the theoretic methodology adopted in the calculations. Key intramolecular interactions within the optimised structures were identified and used to rationalise the energetic ordering. The RESP charges obtained in this work form a comprehensive library of charges for the PIP headgroups, which serve as the crucial groundwork enabling molecular dynamics simulation of PIPs with AMBER.

# Table of Contents

# 1. Introduction

## 1.1 Introduction to membrane lipids

Cell membranes consist primarily of lipids and are crucial for the integrity of cellular structure.[1] Cell membranes also regulate the transport of substances into and out of the cell and modulate the activity of embedded ion-channels and proteins. Although diverse in their structures and functions, all membranes consist of sheet-like noncovalent assemblies of lipids and proteins, **Figure** 1.[2]

Lipids are amphipathic molecules with a hydrophilic headgroup and hydrophobic tailgroup, **Figure 2**. Besides being indispensable membrane components, they serve a variety of biological roles in energy storage, signal transduction, and molecular recognition processes.[4]

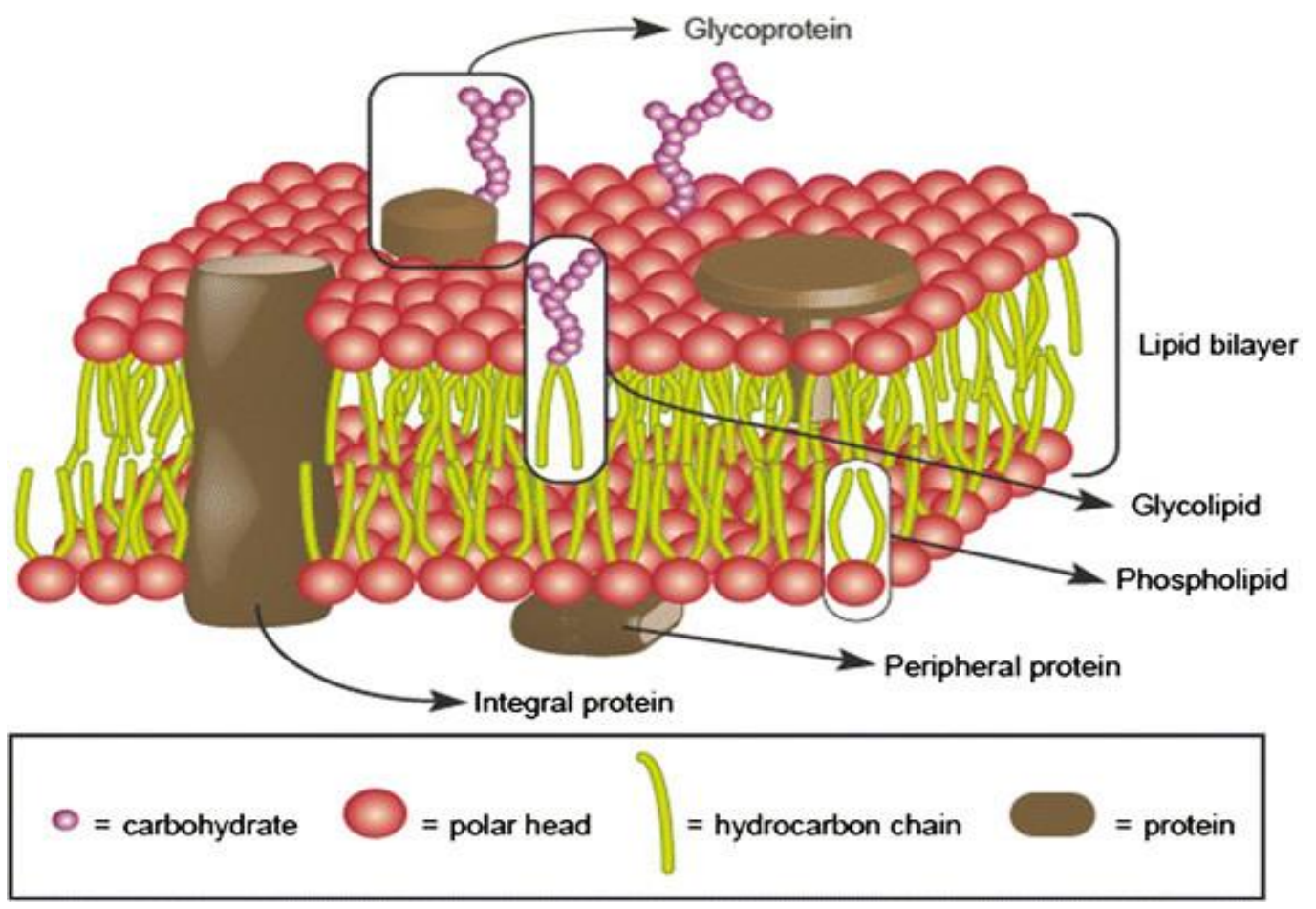

Figure 1: Structure of a biological membrane.[2]

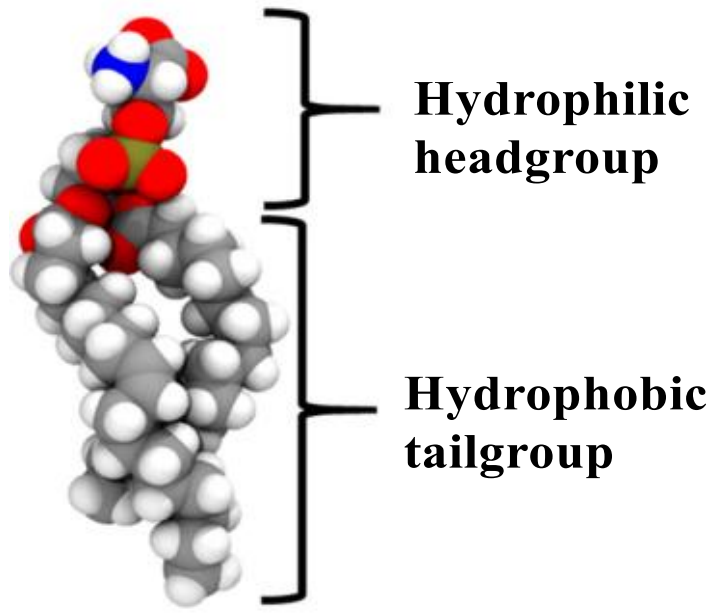

Figure 2: The headgroup and tailgroup in a palmitoyl-oleoyl phosphatidylserine (POPS) lipid molecule.[3]

Lipid bilayer self-assembly is driven by the hydrophobic effect, resulting in molecular aggregates.[5] The balance of geometric factors, environmental conditions and lipid composition dictates that the lipid bilayers are the favoured structures in biological systems, rather than other aggregates such as micelles or inverted hexagonal phases.[6] In a lipid bilayer, the hydrocarbon chains are facing inwards while the hydrophilic headgroups are facing outwards to interact with the aqueous environment.[5]

The three major types of membrane lipids are phospholipids, glycolipids, and cholesterol.[1] Phospholipids are abundant in all biological membranes and serve as a major structural element due to their ability to form bilayer vesicles spontaneously when dispersed in water.[7] A phospholipid molecule consists of a phosphate-containing polar headgroup attached to nonpolar hydrocarbon chain(s). Phospholipids are constructed from four components: fatty acid(s), a phosphate, an alcohol, and a platform (either glycerol or sphingosine, **Figure 3**) to attach the other components onto, **Figure 4**.

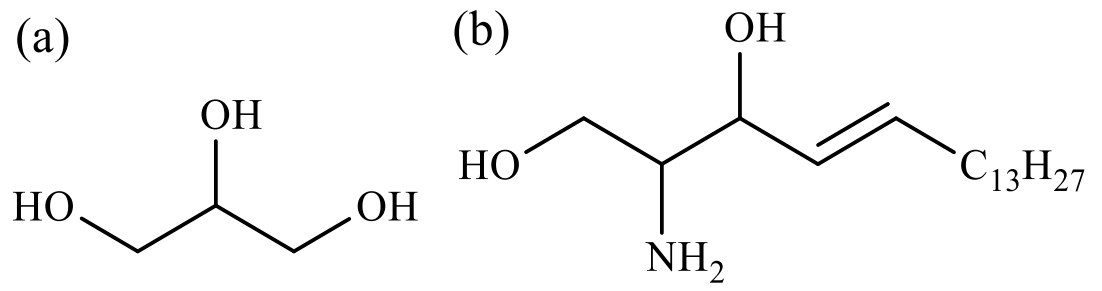

Figure 3: Structures of (a) glycerol or (b) sphingosine.

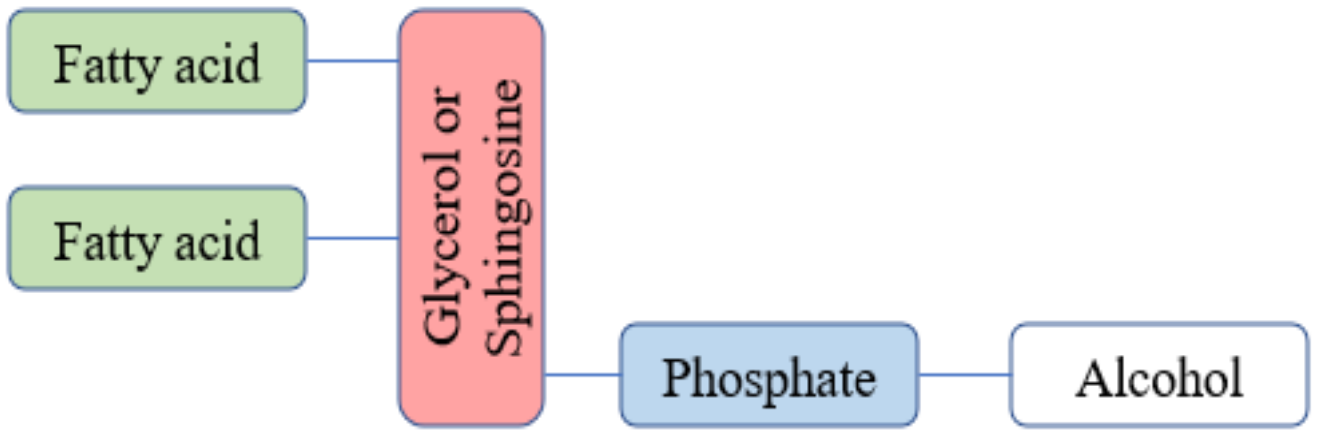

Figure 4: The schematic structure of a phospholipid.

A phospholipid derived from glycerol is called a phosphoglyceride, which consists of a glycerol backbone attached to two fatty acid chains and a phosphorylated alcohol.[1] Phosphoglycerides are derived from the formation of an ester bond between the phosphate group of phosphatidate (**Figure** 5**)** and the hydroxyl group of one of several alcohols. The common alcohol moieties of phosphoglycerides are serine, ethanolamine, choline, glycerol, and inositol, **Figure 6**.[1]

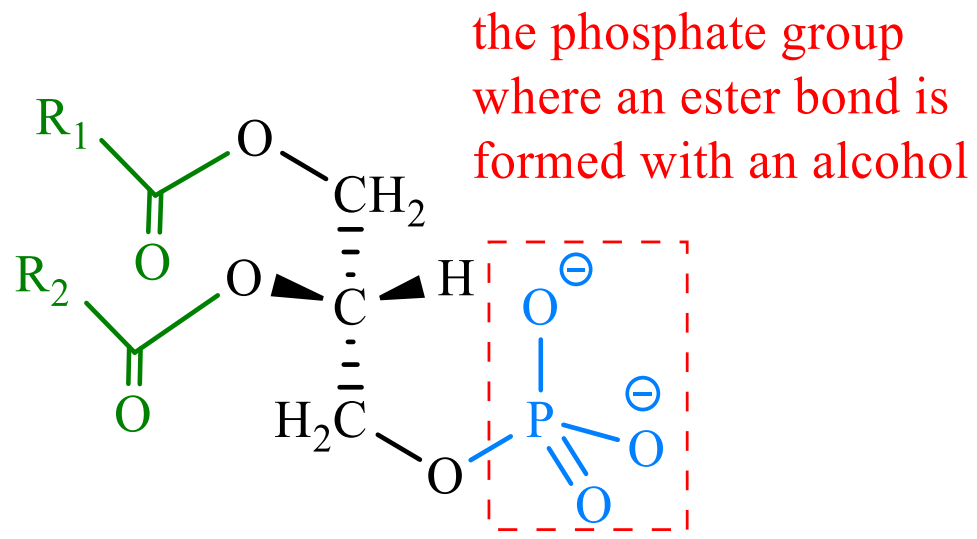

Figure 5: Structure of phosphatidate.

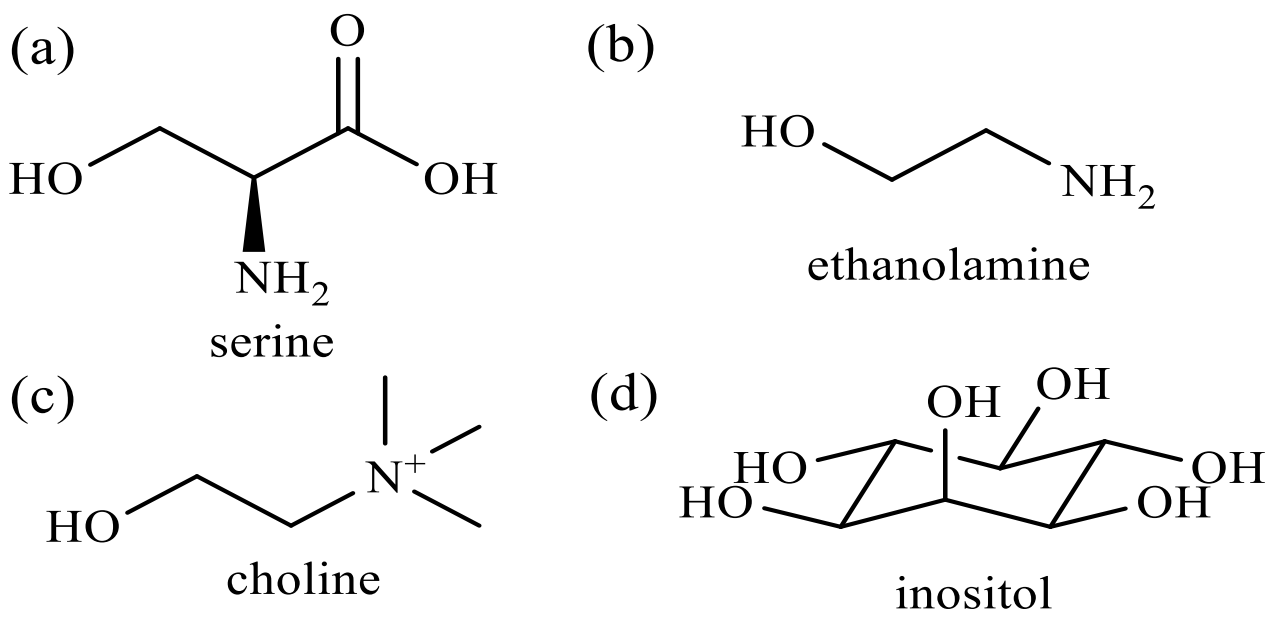

Figure 6: Structures of (a) serine, (b) ethanolamine, (c) choline, and (d) inositol.

## 1.2 Phosphatidylinositol phosphates

Phosphatidylinositols (PIs) consist of a family of lipids in which the alcohol linked to phosphatidate is inositol, **Figure 7**.[1] PIs can be phosphorylated on the 3-, 4- and 5-hydroxyl groups to form a range of mono-, bis-, and tris-phosphorylated products, called phosphatidylinositol phosphates (PIPs), **Figure 8**. PIPs have two hydrocarbon tails, arachidonate and stearate, which are connected through a glycerol group to the inositol headgroup.[1]

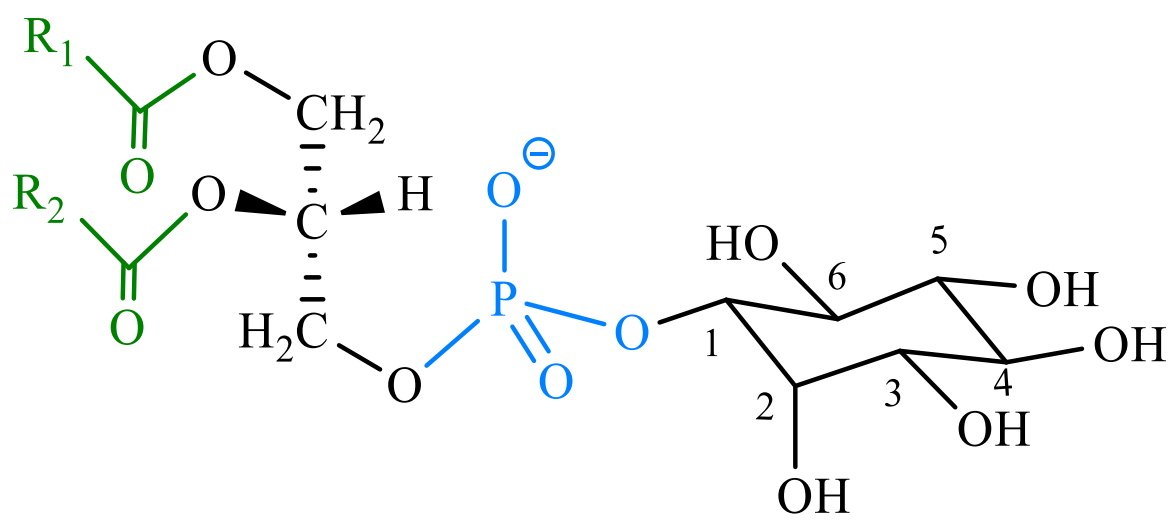

Figure 7: An illustrative structure of PIs.

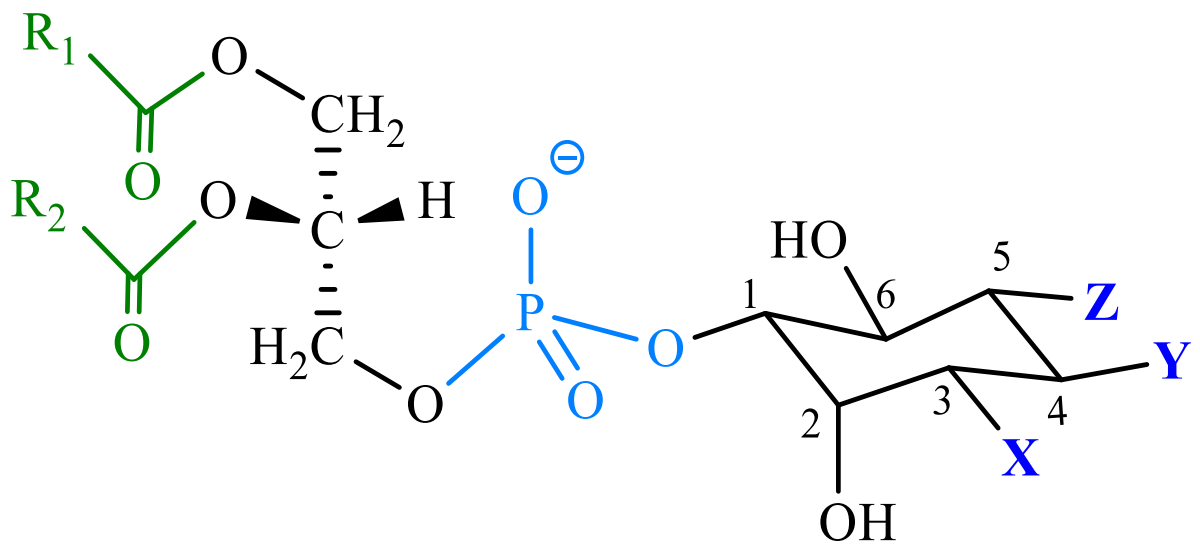

Figure 8: An illustrative structure of PIPs, where the X, Y, Z groups are either OH or $PO_4^{2-}$.

The 7 naturally occurring PIPs differ in the number and position of phosphorylation, with names and structures found in **Table 1** and **Figure 9** respectively.

Table 1: The names of the 7 naturally occurring PIPs with variations in number and position of phosphorylation.

| | X = | Y = | Z = | Generic Names |
|---|---|---|---|---|
| Phosphatidylinositol monophosphates (PIP) | $PO_4^{2-}$ | OH | OH | Phosphatidylinositol 3−phosphate or PI(3)P |
| | OH | $PO_4^{2-}$ | OH | Phosphatidylinositol 4−phosphate or PI(4)P |
| | OH | OH | $PO_4^{2-}$ | Phosphatidylinositol 5−phosphate or PI(5)P |
| Phosphatidylinositol bisphosphates ($PIP_2$) | $PO_4^{2-}$ | $PO_4^{2-}$ | OH | Phosphatidylinositol 3,4−bisphosphate or $PI(3,4)P_2$ |
| | $PO_4^{2-}$ | OH | $PO_4^{2-}$ | Phosphatidylinositol 3,5−bisphosphate or $PI(3,5)P_2$ |
| | OH | $PO_4^{2-}$ | $PO_4^{2-}$ | Phosphatidylinositol 4,5−bisphosphate or $PI(4,5)P_2$ |
| Phosphatidylinositol trisphosphate ($PIP_3$) | $PO_4^{2-}$ | $PO_4^{2-}$ | $PO_4^{2-}$ | Phosphatidylinositol 3,4,5−trisphosphate or $PI(3,4,5)P_3$ |

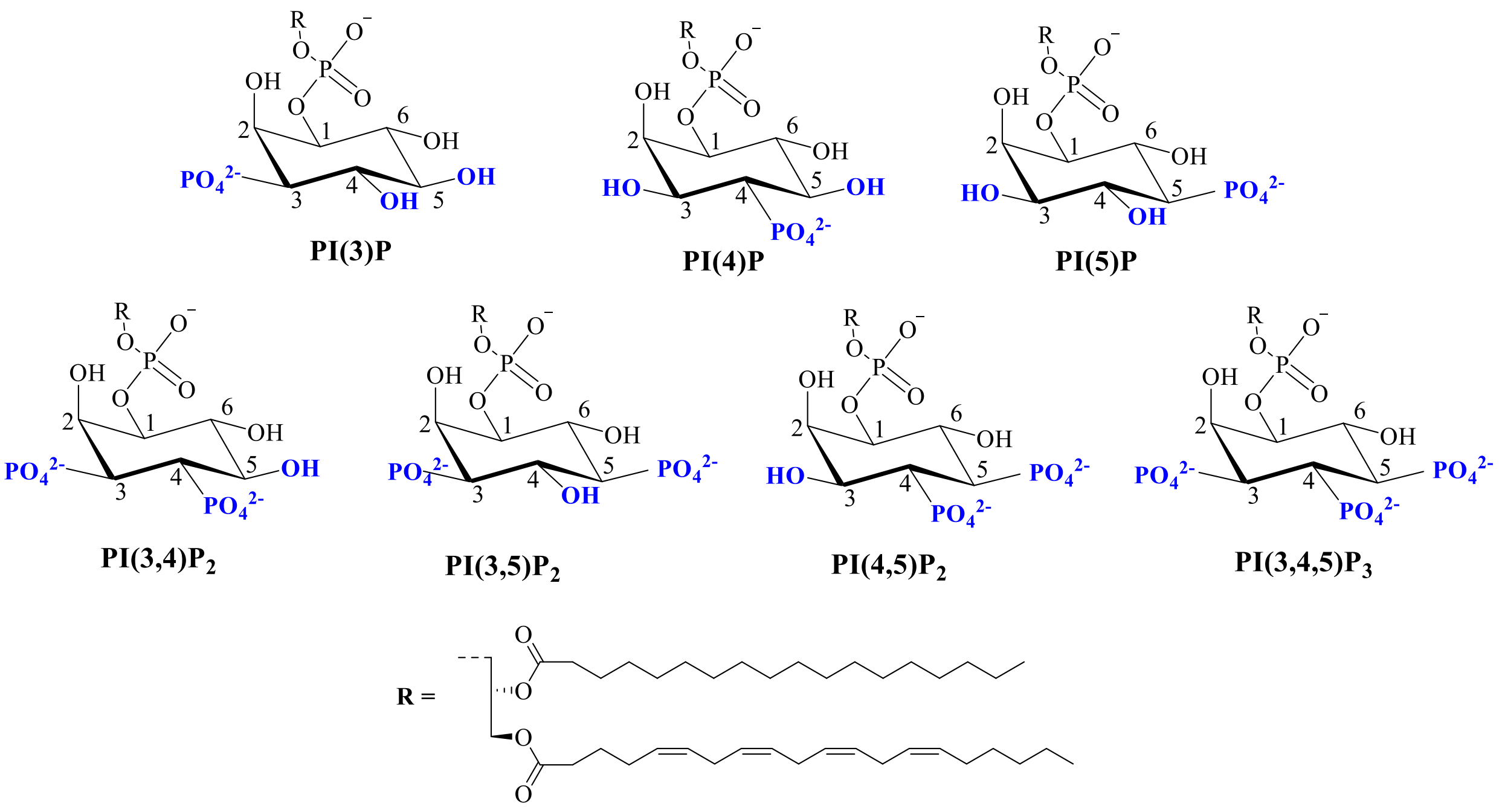

Figure 9: The structures of the 7 naturally occurring PIPs, differing in the number and position of phosphorylation. The phosphate groups shown are fully deprotonated.

### 1.2.1 Biological roles of PIPs

Although PIPs only play a minor structural role in lipid bilayers, constituting 1–2% of the total phospholipids in the plasma membrane, they are indispensable in regulating cellular behaviours.[8] The distinct arrangements of hydroxyl and phosphomonoester substitutions on the inositol ring are responsible for the specificity of many PIP-protein interactions.[9]

The principal role PIPs serve is to interact with proteins, through which they select peripheral proteins for membranes and regulate the activity of integral proteins.[10] For example, PI(3)P plays a key role in regulating membrane trafficking, which distributes proteins and other macromolecules throughout the cell.[11] PIPs also act as substrates for a variety of PIP-effector proteins. The binding is usually highly selective through specialised domains on the proteins.[10] For instance, ion channels are integral proteins in the plasma membrane that often require binding to the plasma membrane-specific PI(4,5)$P_2$ to function.[12]

PIPs also play crucial roles in numerous intracellular signalling pathways.[12] PI(3,4,5)$P_3$ is an important second messenger in mammalian cells that binds to the protein kinase AKT to promote cell survival.[13] PIP turnover inside the cell is tightly controlled by metabolic regulatory enzymes such as specialised kinases, and defects in the metabolism are associated with cancer, cardiovascular disease, and autoimmune dysfunction.[14]

### 1.2.2 Charged states

The charge states of the phosphomonoesters in PIPs varies depending on the pH. At physiological pH, the phosphomonoesters are usually negatively charged and can adopt either a fully deprotonated or singly deprotonated state, **Figure 10**.[15] This results in a number of variations of the charge states for phosphatidylinositol mono-, bis- and tris-phosphates, **Figure 11.**

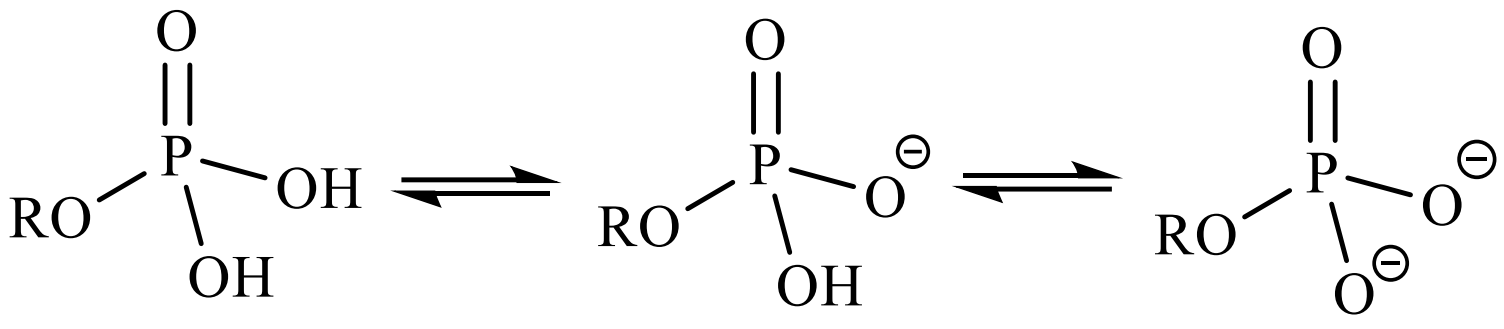

Figure 10: Different ionisation/charge states of phosphomonoesters.

As it is difficult to assign the exact charge states of the phosphomonoesters in different PIP lipids due to their extremely complex ionisation behaviours, many studies simplify the problem by only considering the fully deprotonated states.[815] However, *in vitro* studies suggest that full deprotonation can be highly unlikely or even unfavourable in some cases, as full deprotonation will lead to high charges concentrated on a small molecule.[16] Especially for trisphosphates, where full protonation can leads to a −7 charge concentrated on the PIP headgroup.

As many roles and properties of PIPs are dependent on their ionisation states, all 26 possible charge states of the 7 biologically-active PIPs were investigated in a systematic approach as illustrated in **Figure 11**.

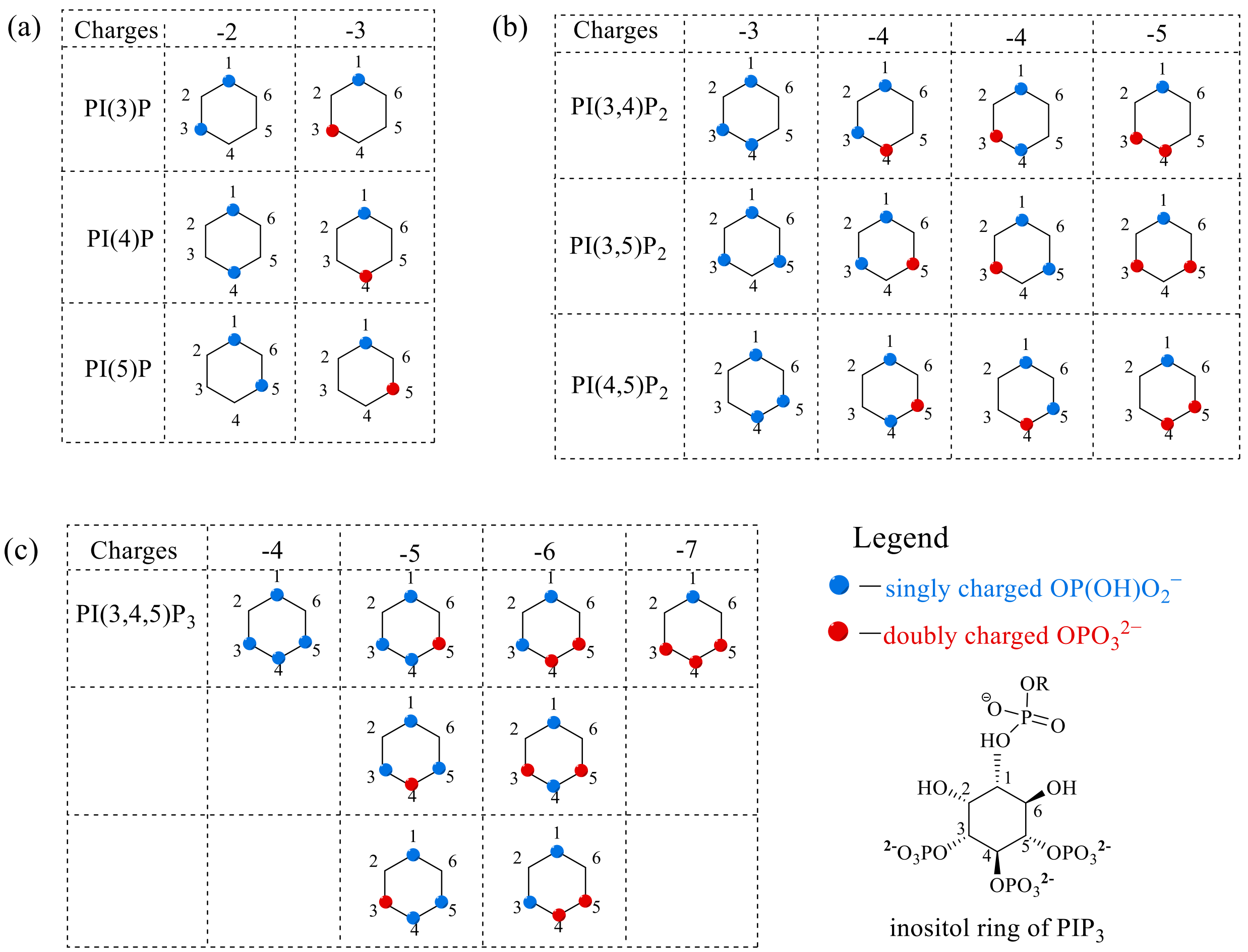

Figure 11: Schematic illustrations of the 26 possible charge states for phosphatidylinositol (a) mono-, (b) bis- and (c) tris-phosphates. The hexagons represent the inositol rings in PIPs, while the blue and red circles represent the singly and doubly deprotonated phosphate groups.

## 1.3 Methods to study lipid membranes

### 1.3.1 Experimental methods

Understanding membrane structure and transition helps to unveil the complexity in biological systems and improve our ability to model biology. It can also provide inspirations to new soft materials or novel drug targets.[17]

The primary methods to probe membrane structural properties are X-ray crystallography, electron microscopy, and NMR spectroscopy.[18] Neutron scattering and calorimetric methods can also be used. These complementary methods have contributed significantly in our currently understanding of membrane structures and functions.

However, experimental investigations can be time-consuming and expensive, and the feasibility can be highly dependent on the equipment and skills available. For instances, X-ray crystallography requires the growth of well-ordered single crystals and the use of highly monochromatic and collimated X-rays.[18] To resolve the dynamics in lipid phase transitions, extremely fast and sensitive probes are needed. Moreover, some lipid samples are extremely difficult to isolated and might easily decompose, which further limit the practicality of experimental studies in some cases.[18]

### 1.3.2 Limited PIP experimental data

Due to the crucial roles PIPs play in a wide range of biological functions, the structures and properties of PIP-containing membranes are of intense research interest. However, there is a dearth of experimental data for PIPs in the literature.[19] As a result, many PIP-protein interacting domains and stereoselectivities remain elusive.[20]

The scarcity of experimental results could be mainly attributed to the prohibitively-expensive nature of PIP research. Not only is the use of analytical techniques such as HPLC or X-ray crystallography logistically challenging, time-consuming, and costly,[21] but commercial PIP compounds also come with a rather steep price tag (hundreds of dollars per mg). The high prices for commercial PIPs arise from the complex synthetic strategies and purification procedures,[22] which make growing large crystals for X-ray crystallography or producing isotope-enriched samples for NMR extremely expensive.[23] Hence, there is considerable motivation in exploring alternative techniques to investigate the structures and properties of PIPs.

### 1.3.3 Computational methods

With the rapid improvements in computing capacity and reduction in computing cost, many *in silico* methods have been developed to study membrane properties.[8] Computational simulation of membranes can be very versatile, providing atomic level resolution and theoretical support to *in vivo* experiments. It could also be used to shed light on systems where experimental results are lacking or unavailable.

Data extracted from sophisticated computer modelling and simulations are aligning increasingly close to available experimental data, thus can be used to give useful results that are not easily measurable through experimental work.[24] Due to the lower cost both in terms of time and resource, computational studies of membrane systems have been gaining rapid popularity.[25] These *in silico* methods can be used as attractive supplements or even possible alternatives to experimental studies of PIP-containing lipid membranes.[8]

## 1.4 Quantum chemical methods

*Ab initio* quantum chemical (QC) methods are often used to investigate the chemical properties, transition state structures, and reaction mechanisms that are experimentally difficult to examine.[26] QC methods calculate the energy and electronic structure of a molecular system by solving the Schrodinger equation, **Eq 1**.[27] Through *QC* methods, the structural, electronic, vibrational and NMR properties can be obtained.[28]

$$\hat{H}\Psi = E\Psi \qquad (1)$$

A pioneering QC method is the Hartree-Fock (HF) method, which models the motion of electrons independently as described in the Born-Oppenheimer approximation. The wavefunction and the energy of a many-body quantum system in stationary state are determined.[29] The HF method approximates the electronic wavefunction of a N-body system with a single Slater determinant of N spin-orbitals:[29]

$$\Psi(1,\ldots,N) = \frac{1}{\sqrt{N!}}\begin{vmatrix} \psi_1(1) & \psi_1(2) & \cdots & \psi_1(N) \\ \psi_2(1) & \psi_2(2) & \cdots & \psi_2(N) \\ \vdots & \vdots & \ddots & \vdots \\ \psi_N(1) & \psi_N(2) & \cdots & \psi_N(N) \end{vmatrix} \qquad (2)$$

*Ψ:* wavefunction of 4*N* degrees of freedom

*N:* no. electrons

A set of N-coupled equations for N spin-orbitals can then be derived using the variational method,.[30] A solution to these equations solves the Schrodinger's equation and yields the HF energy of the system as well as the wavefunction.[29] However, as electron-correlation are not accounted for because the electron-electron interactions are included as a mean field, the resulting energy will be higher than the exact solution.[30]

### 1.4.1 MP2

The Møller-Plesset perturbation theory (MP) is a post-HF *ab initio* QC method that improves on the HF method by adding electron correlation effects through Rayleigh-Schrödinger perturbation theory (RS-PT),[31] commonly to the second (MP2), third (MP3) or fourth (MP4) order. However, this treatment is associated with significant increase in computational cost.[32]

In particular, MP2 is one of the simplest and most useful level of theory beyond the HF approximation[31] as it offers systematic improvement in optimised geometries and other molecular properties relative to HF theory.[32] Compared to the more economical density functional theory (DFT) methods,[33] MP2 also has the advantage of properly incorporating long-range dispersion forces.[34]

### 1.4.2 Basis sets

A basis set is a set of wavefunctions that describes the shape of atomic orbitals (AOs) and are required in virtually all *ab initio* calculations of molecular systems.[35] The accuracy of the approximations used in QC calculations are directly related to the basis sets used, while computational cost shows an inverse relation.[36] The optimal basis set should describe all parts of the molecule well and allow realistic calculations of the desired properties, while being as small as possible to minimise computational costs. Hence, the choice of basis sets often faces a trade-off between the accuracy of results and CPU time.[35]

#### 6-31G*

The 6-31G* basis set is a popular split-valence basis sets derived from 6-31G basis set augmented by diffuse and polarisation functions.[35] 6-31G* basis set is less robust than higher-level ones (such as the Dunning-Huzinaga basis sets),[37] but the cost is substantially lower both in terms of computational time and resource (CPU, memory, and disk).[38] However, the overall discrepancy between experimental values and those calculated with the 6-31G* set can be rather significant.[35] Hence, in this study 6-31G* is only used in the initial stage of the optimisation process in order to minimise the optimisation steps involved, before higher basis sets are used.

#### Correlation-consistent basis sets

Correlation consistent (cc) basis sets can accurately describe core-core and core-valence correlation effects.[39] For first and second row atoms, the basis sets cc-pVNZ are used, where N = D (double), T (triples), Q (quartet), which successively contain larger shells of polarisation functions (d, f, g). The 'cc-p' stands for 'correlation-consistent polarized' and the 'V' indicates they are valence-only basis sets.[40] As correlated (post-HF) calculations are more accurate, the QC calculations in this study are subsequently performed with cc-pVDZ and finally with cc-pVTZ basis sets.

### 1.4.3 Polarizable Continuum Model

Solvation effects can be included in QC calculation using explicit or implicit solvent models,[41] where the solvents are modelled by a large number of explicitly modelled solvent molecules surrounding the solutes,[42] or by a continuous and homogeneous solvent medium,[43] respectively. Implicit models are generally less computationally expensive because an averaged dielectric continuum is used instead of explicit solvent molecules.

There are a number of  commonly used implicit solvent models, which differ in their description of the solute-solute interactions.[44] In this project, the polarizable continuum models (PCM) is applied in the QC geometry optimisation calculations. As the name suggests, PCM models the solvent as a polarisable continuum and mainly include electrostatic solute-solvent interactions.[45] It surrounds the solute with a solvent cavity by placing charges (modulated by the dielectric constant of the solvent) on the surface of the cavity, so as to mimic the charge stabilisation of the solvent environment.[46] Qualitatively, the solute inside the cavity experiences the effect of the solvent environment created by charges at the cavity boundary, **Figure 12**.[47]

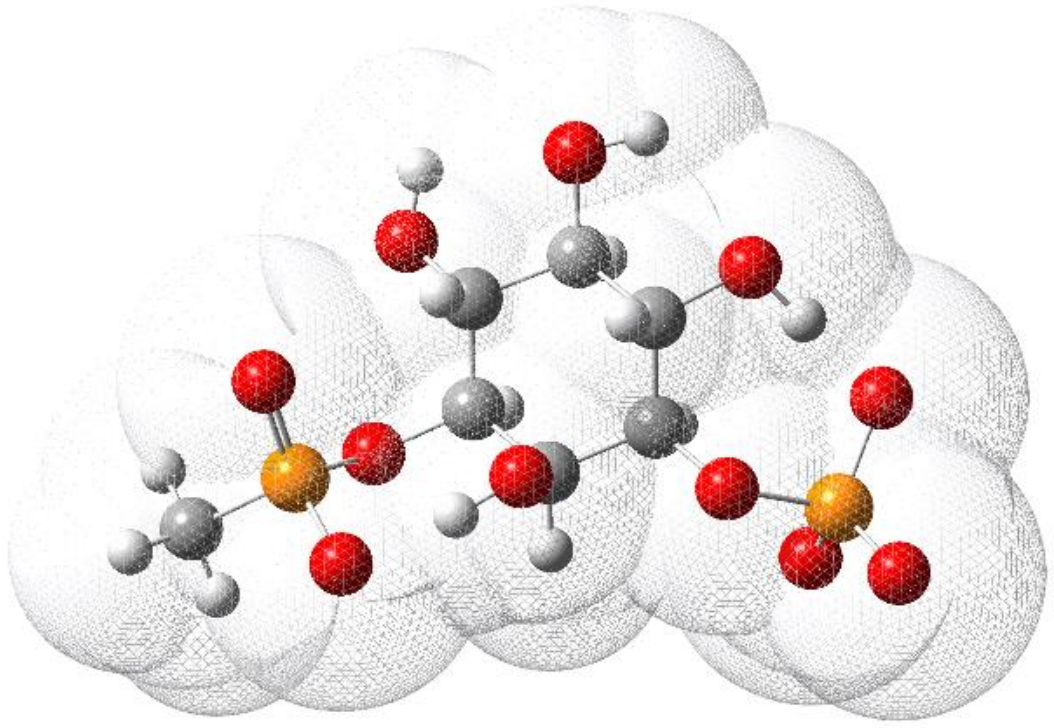

Figure 12: An illustration of PCM solvent cavity with PI(3)P (−3 charged) as the solute and water as the solvent.

In this work, all QC calculations for the PIP headgroups were performed in implicit water solvent using PCM. This treatment accounts for the electrostatic interactions between PIPs and water, which are important for the calculations conducted in this project. This is because the highly negatively-charged PIP headgroups are unstable when isolated in the gas phase, especially for the fully deprotonated states. Hence, water molecules are needed to stabilise the charges. Moreover, as PIP molecules are of biological significance, conducting calculations in a water environment will allow for comparison of the theoretical results obtained from the calculations with the available *in vitro* data.

Although QC methods can be very versatile in investigating structural and electronic properties, they are too computationally intensive to be used for large systems comprised of many atoms such as lipid membranes.[27] Nevertheless, QC methods can be extremely useful in deriving parameters for individual lipid molecules, which can then be used in less computationally demanding methods to investigate membrane properties.

## 1.5 Molecular Mechanics

Molecular mechanics (MM) is a computational method that uses classical physics to study the properties of molecular systems.[48] MM is particularly useful for studying the time dependent properties of large systems, where quantum mechanical (QM) treatment is too computationally expensive.[49]

MM treats each atom in a system as a hard, charged sphere connected to others through a spring. The interactions between the atoms (potential energy) are described mathematically with empirical parameters, derived either experimentally or computationally.[48] As MM methods ignore electronic motions and calculate the energy of the system as a function of the nuclear positions only, they have significantly lower computational cost compared to QC methods, thus enabling MM to be used on larger systems.[50]

### 1.5.1 Force fields

A force field is a mathematical function which describes the potential energy of a system. MM force fields approximate the QM energy surface with a classical mechanical model and are the method of choice for lipid simulations.[48]

Typical force fields have a potential energy function, $E_{\text{potential}}$ that includes the following bonded and non-bonded terms, **Eq 3**.

$$\begin{aligned} E_{\text{potential}} &= E_{\text{bonded}} + E_{\text{non-bonded}} \\ &= E_{\text{bond}} + E_{\text{angle}} + E_{\text{torsion}} + E_{\text{elec}} + E_{vdw} \end{aligned} \tag{3}$$

$E_{bond}$ = bond stretching and bending energies

$E_{\text{angle}}$ = angle energies

$E_{torsion}$ = dihedral or torsional energies

$E_{elec}$ = electrostatic interactions

$E_{vdw}$ = Van der Waals interactions

## 1.6 Molecular Dynamics

Molecular dynamics (MD) uses MM-based potential energy functions to study the time dependent behaviour of a molecular system. MD simulations generate successive configurations of a system over time, calculating the forces acting on the atoms and updating their positions based on those forces. The force acting on the atom is calculated from the first derivative of the potential energy function:

$$F = -m\frac{dV}{dr} \tag{4}$$

The net force on each atom in the system is the sum of forces from all other atoms in the system:

$$\vec{F_i} = \sum_{j=1}^{n_{max}} f_j, (j \neq i) \quad (5)$$

For a system comprised of a collection of N atoms, each atom experiences a force that causes it to accelerate, given by the Newton's second law of motion:

$$F_i = m_i a_i = m_i \frac{dv_i}{dt} = m_i \frac{d^2 x_i}{dt^2} \quad (6)$$

$F_i$ = force acting on atom *i*

$m_i$ = mass of atom *i*.

$a_i$ = acceleration of atom *i*

$v_i$ = velocity of atom *i*

$x_i$ = position of atom *i*

This second order differential equation allows the determination of the positions and velocities of each atom when the force $F_i$ is known.[51] This is usually solved through numerical integration methods, such as the commonly used Classical Verlet algorithm (**Eq 7**) and the Velocity-Verlet algorithm (**Eq 8**):[52]

$$x(t + \delta t) \approx 2x_i(t) - x_i(t - \delta t) + \frac{F_i(t)}{m_i}\delta t^2 \quad (7)$$

$$\mathrm{v}_i(t + \delta t) = \mathrm{v}_i\left(t + \frac{1}{2}\delta t\right) + \frac{1}{2}\mathrm{a}_i(t + \delta t)\delta t \quad (8)$$

The output of MD simulation is a trajectory which contains atomic positions in time, which is obtained through following the dynamics of a single system.[53] A time average over a series of measurements (atomic positions and velocities) gives the conformational ensemble profile, which is a large collection of copies of a system, each of which represents a possible state that the real system could be in.[54, 55]

Simulating a large number of replicas of the system simultaneously to obtain the ensemble average can be extremely computationally expensive,[56] hence the trajectory of a single system over time is simulated in MD. This treatment is based on the ergodic hypothesis, which states that time average equals the ensemble average.[57]

The central idea of the ergodic hypothesis is that a system allowed to evolve in time indefinitely should eventually pass through all possible states for the particular thermodynamic constraints.[58] Therefore, MD simulations need to sample a sufficient amount of phase space to generate enough representative conformations so that the Ergodic hypothesis is satisfied. Experimentally relevant information concerning structural, dynamic, and thermodynamic properties can then be calculated from the time-averaged microscopic information through statistical mechanics.[57]

## 1.7 MD simulation of lipid bilayers

MD is a valuable technique which enables the investigation of statistical and dynamic properties of a real poly-atomic system at the molecular level.[54] In the context of lipid molecules, MD simulation can facilitate the studies of numerous membrane phenomena, such as bilayer phase transitions, vesicle dynamics, and the behaviour of realistic cell membranes. MD simulations of biological molecules such as lipids, proteins, and nucleic acids are challenging due to the distinct and complicated bonded and non-bonded interactions that are involved in the systems.[25]

Over the years, a suite of software packages have been developed for biochemical systems, such as GROMACS,[59] CHARMM,[60] and AMBER.[61] Many chemical-specific lipid force fields (FFs) have also been developed with varying levels of resolutions, and can be broadly grouped into three types: all-atom (AA), united-atom (UA), and coarse-grain (CG), **Figure 13**.[3] When detailed studies of membrane properties are required, AA and UA are usually used. In particular, AA models are used when examining key membrane properties that require atomistic details, such as hydrogen bonding[62] and lipid NMR order parameters.[63]

The validity of results obtained using MD methods depends largely on the FF that is used. An accurate lipid FF must be able to capture the balance of the chemical-specific interactions in the system simulated to represent phase changes. It must also aptly represent the internal molecular structures important for describing the bond distances, angles, and dihedral states along with the intermolecular interactions important for forming self-assembled structures. In order to facilitate meaningful *in silico* studies of lipid membranes, lipid FFs must be parameterised appropriately, so that results from simulations are in agreement with available experimental results from literature.[64]

### 1.7.1 AMBER and Lipid force fields

AMBER is a widely-used MD package that contains a collection of programs and FFs primarily designed for the simulation of biomolecules.[61, 65] AMBER supports a wide range of parametrised AA FFs covering proteins and nucleic acids (ff12SB),[66] carbohydrates (Glycam),[67,68] and lipids (Lipid14),[69] as well as general organic molecules (GAFF).[70] The FFs in Amber have the general form as shown in **Eq 9**, and are parametrised to suit each specific system.

$$E_{total} = \sum_{bonds} \frac{k_l}{2}(l - l_0)^2 + \sum_{angles} \frac{k_\theta}{2}(\theta - \theta_0)^2 + \sum_{dihedrals} \frac{V_N}{2}[1 + \cos(n\omega - \gamma)] + \sum_{i=1}^{N} \sum_{j=i+1}^{N} \left(4\varepsilon_{ij}\left[\left(\frac{\sigma_{ij}}{r_{ij}}\right)^{12} - \left(\frac{\sigma_{ij}}{r_{ij}}\right)^{6}\right] + \frac{q_i q_j}{4\pi\varepsilon_0 r_{ij}}\right) \quad (9)$$

The AMBER Lipid FF is characterised by its modular nature which allows for numerous combinations of head and tail groups to create a variety of lipid types, **Figure 14**. The revisions and updates of Lipid FF parameters are a continuous progress. Although much progress have been made in the Lipid FF development over the years, many aspects can still be revised to improve the accuracy of simulated results.[71] Examples include poor agreement between simulated and experimental NMR head group order parameters and mismatched phase transition temperatures.[72] There is also scope for extending Lipid FFs to include a wider selection of lipid headgroups and chains, as diversity in the parameterised lipid components are crucial for the development of a versatile lipid FF that can model realistic biological membranes.[3]

PIPs are important components in many biological membranes. However, the parameterisation of PIPs has been lagging behind other membrane lipids. Only in recently years PIPs parameters have been gradually incorporated in selected lipid FFs, such as in GROMOS and MARTINI.[3] Many commonly used FFs have yet to parameterise PIPs, including the latest Lipid21 FF used in AMBER.[73]

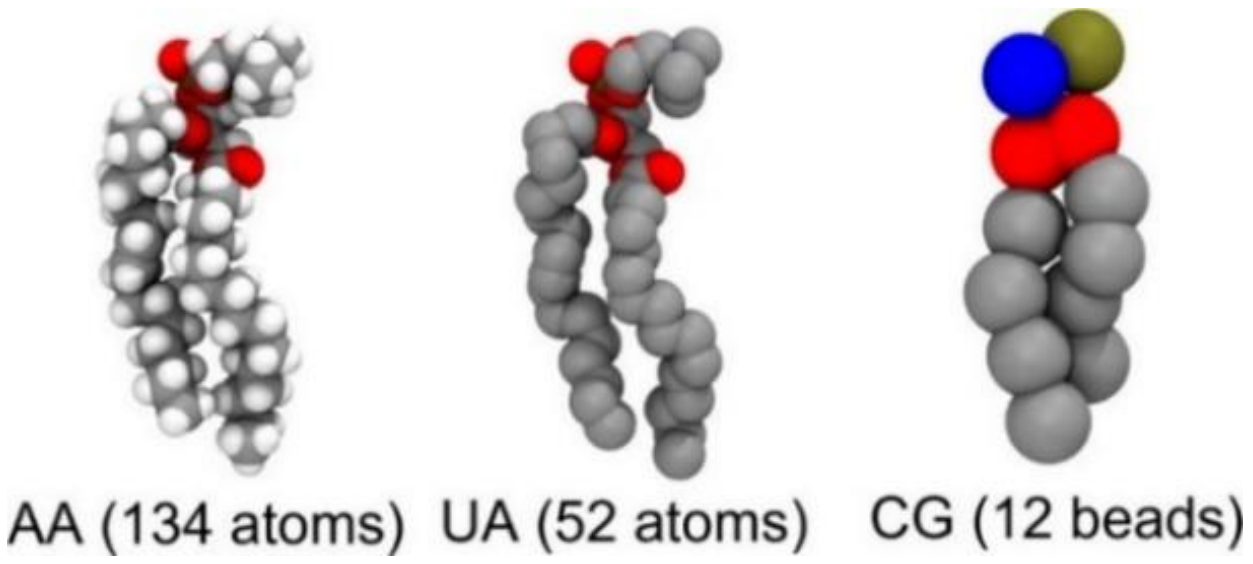

Figure 13: Comparison of AA, UA, and CG force field atoms/particles for a POPC lipid.[3]

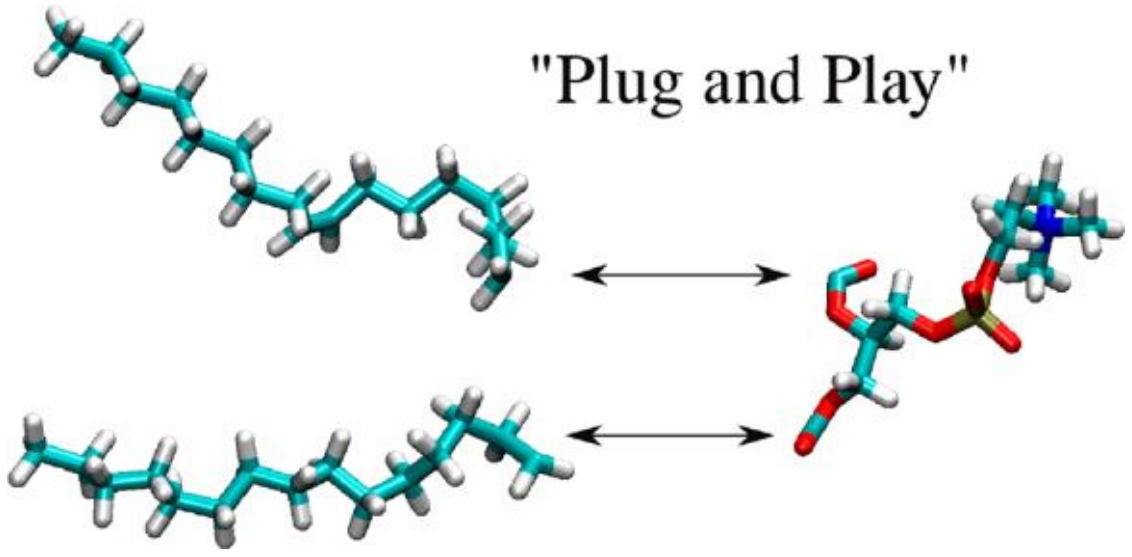

Figure 14: An illustration of the modular nature of the AMBER Lipid force field, where head and tail groups are pluggable and act as building blocks for a variety of lipid types.

# 2. Aims and Objectives

## 2.1 Overall aim

As there are currently no existing parameters for PIP headgroups in the AMBER FFs, the overall aim this research is to extend the latest Lipid21 FF with parameters for the PIP species. The parameters will then be validated by ensuring the lipid bilayer properties derived from MD simulations agree well with available experimental data from literature. The inclusion of well-validated PIP parameters will allow for MD simulation of PIP-containing systems, expanding the types of membrane systems that can be study with AMBER. As PIPs are integral components in many biological membranes, the ability to simulate PIPs with AMBER will enable many crucial structures, properties and interactions of PIPs to be investigated theoretically.

## 2.2 Project objectives

Previous attempts were made within the Gould group to generate charge parameters for PIPs at a relatively low level of theory (HF level), but a complete set of geometries and charges were not obtained due to issues with geometry optimisation as well as the extremely time-consuming nature of these calculations.[74] Hence, the objectives of this project can be summarised as follows:

- To obtain optimised geometries for all 7 biologically-active PIP headgroups with high-level of theory and implicit solvation, in line with the latest Lipid21 FF protocol. With the aim of creating a comprehensive library of PIP charges, every possible charge state for PIPs will be considered.
- To analyse the relation between the calculated energetic ordering of PIPs with the position of phosphorylation and the degree of ionisation. This will be achieved by ranking the calculated energies within each charge subset and rationalising the ordering with the possible stabilising and destabilising interactions present in the optimised geometries. Comparison will also be made between the energetic ordering from the calculations and *in vitro* data available from literature.
- To derive PIP charge parameters compatible with the Lipid21 FF. This will be done by firstly calculating the electrostatic potential (ESP) which are used to derive the restrained electrostatic potential (RESP) charges via a two-stage RESP fitting protocol as applied in Lipid21. The RESP charges reproduce the electrostatic field around the PIP molecules and are well-suited for studying intra- and intermolecular interactions and molecular properties with AMBER.

# 3. Computational Methods

## 3.1 Parameterisation protocol overview

The parameterisation protocol to be used in this work for PIP headgroups is in keeping with the AMBER Lipid21 force field development.[73] A simplified outline is illustrated in **Figure 15**. The work conducted in this project focuses on step 1 and 2.

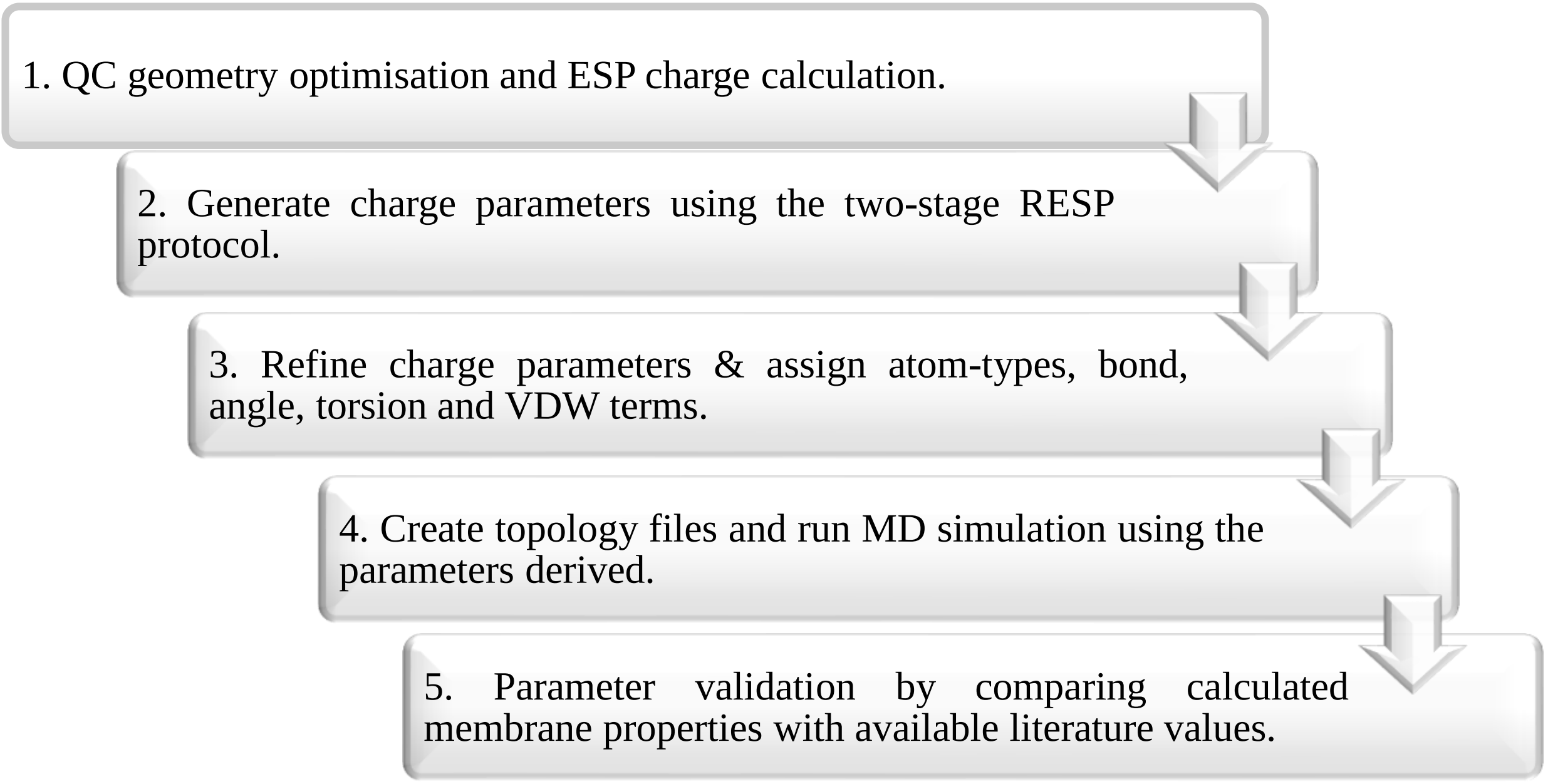

Figure 15: Schematic summary of parameterisation protocol to be used in the study.

## 3.2 Geometry optimisation with Gaussian

QC calculations have been performed with Gaussian 16 (Rev. C.01).[75] The initial geometries of all 26 possible PIP charge states used in the calculations were constructed in GaussView (6.0.16),[76] **Figure 16**. The methyl headgroup capping procedure was applied in accordance with the latest Lipid21 protocol.[73] Each structure underwent optimisation at the MP2/6-31G*, MP2/cc-pVDZ, and MP2/cc-pVTZ level with the Polarizable Continuum Model and water as the solvent. The optimised structures were confirmed as local minima on the potential energy surfaces through checking for convergence.

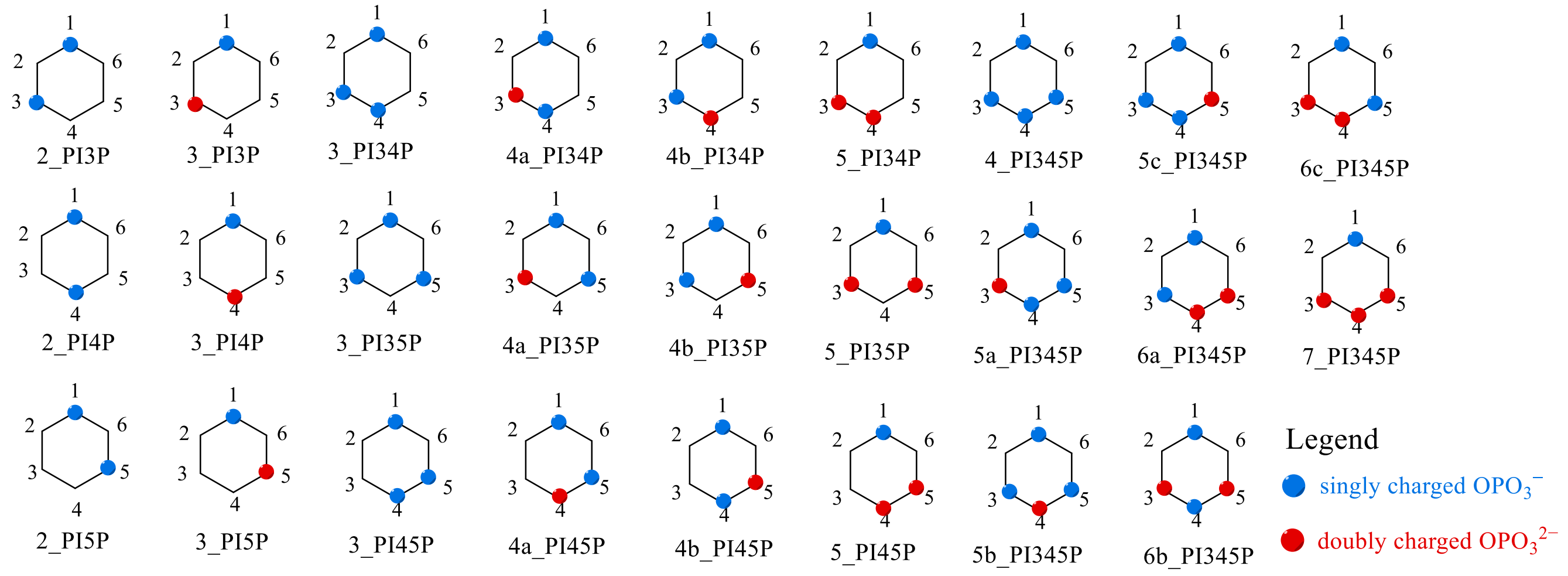

Figure 16: Representations of the 26 PIP geometries used in the QC calculation and their assigned names, where the first number represents the respective net charge on each PIP structure.

## 3.3 Capping strategy

The capping strategy adopted for the PIP headgroups were in line with the Lipid21 protocol.[73] For each PIP structure, the ester group linking the glycerol backbone and fatty acid tails were considered part of the headgroup residues, which is capped with a methyl group, **Figure 17**. These methyl-capped PIP headgroups were the structures used in the QC optimisations and charge derivations.

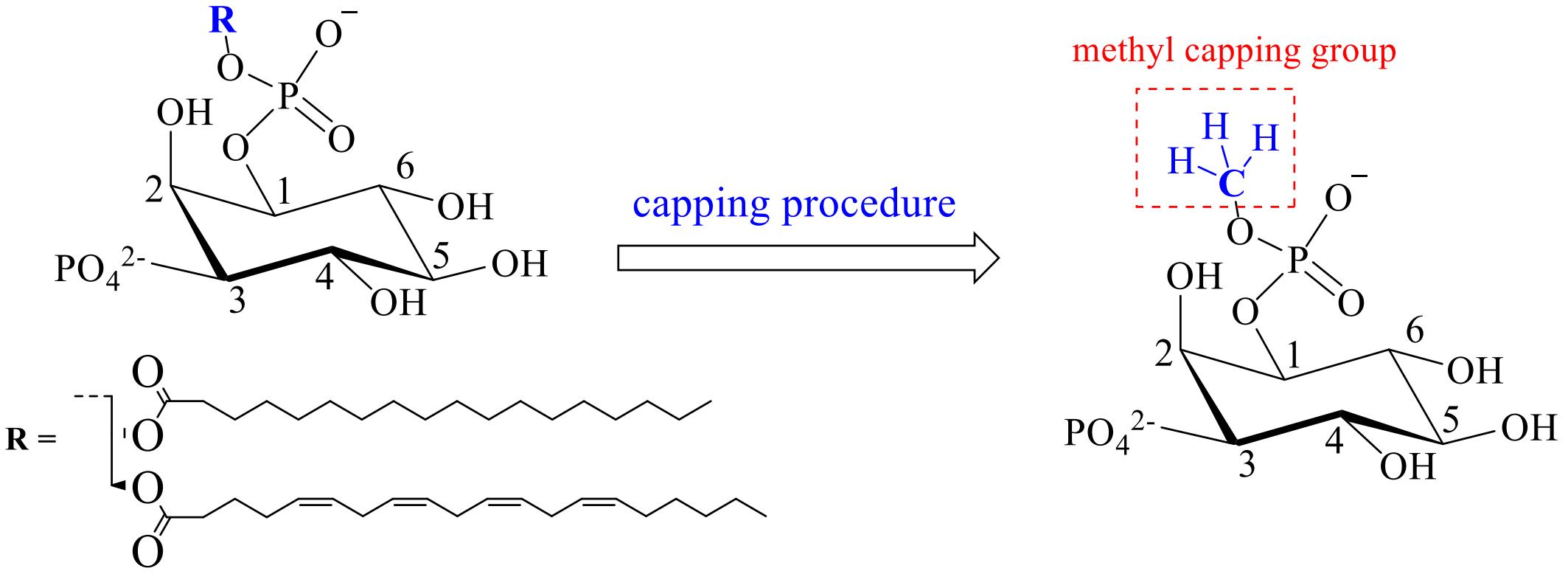

Figure 17: An illustration of the capping procedure for the headgroup of 3_PI3P.

The charges for the capping groups were derived from methyl acetate (headgroup and tailgroup caps bonded together) at the MP2/cc-pVTZ level, with the net charge of each of the capping groups constrained to zero, **Figure 18**. The methyl capping group charges were applied as constraints in the subsequent RESP fitting procedure for all PIP headgroups.

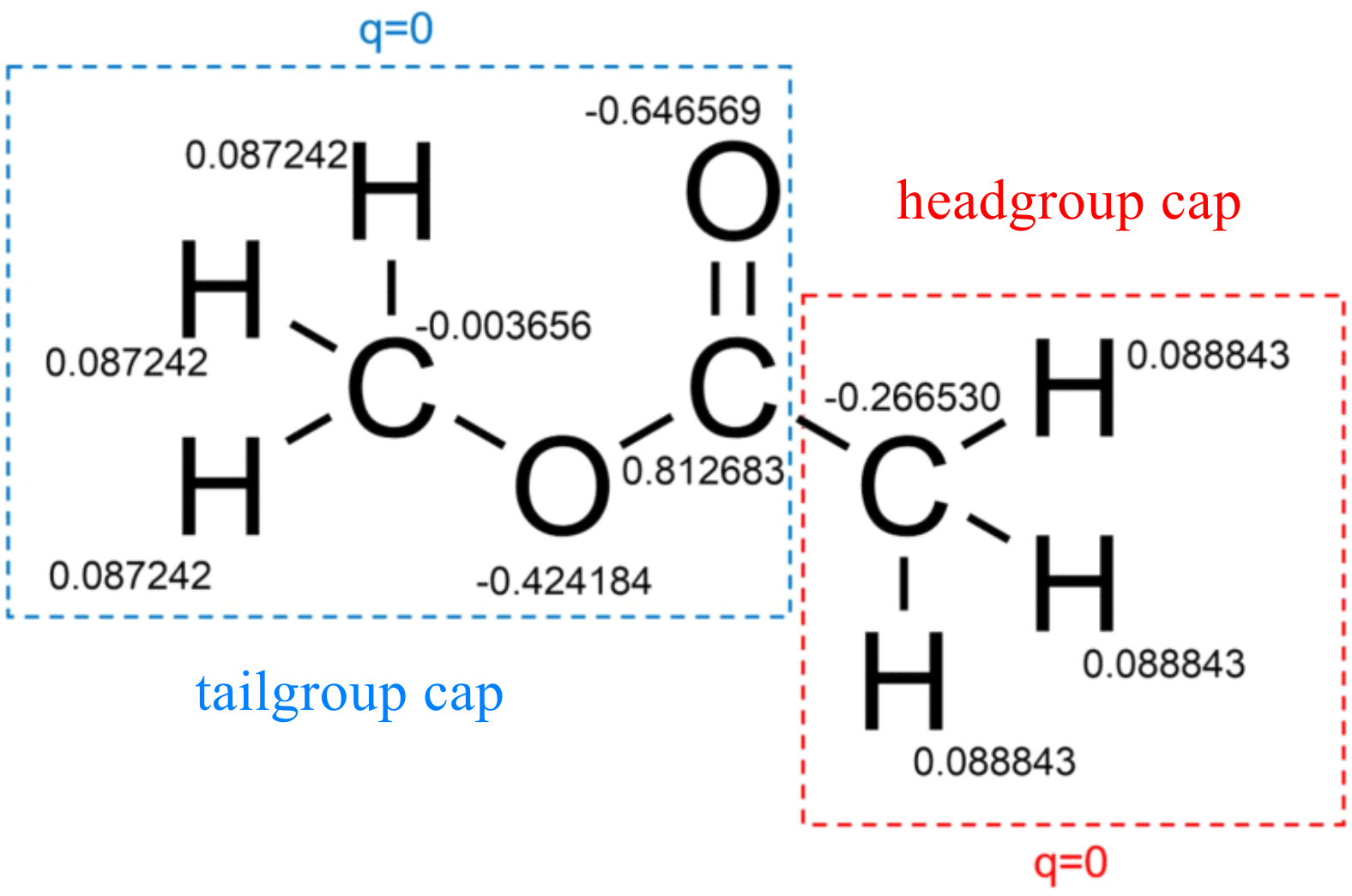

Figure 18: Methyl acetate partial charges for head and tail group caps, derived at MP2/cc-pVTZ level with Polarizable Continuum Model.[73]

### 3.4 ESP calculation and RESP fitting

The molecular ESP was calculated for each optimised PIP structure at the MP2/cc-pVTZ level. Graphical representations of the ESPs were then produced with GaussView.[76] Partial charge parameters were subsequently derived for each of the PIP headgroups using a two-stage RESP fitting procedure[73] to derive atom-centred point charges that best reproduced the ESP. Constraints were applied during RESP fitting to ensure that the methyl capping groups have the correct charges, while maintaining the correct net charges on the PIP headgroups.

# 4. Results and Discussion

All 26 Gaussian optimisation jobs for the different PIP headgroup structures had successfully converged at the MP2/cc-pVTZ level. Multiple attempts were required for some jobs to converge due to the high memory and disk requirements associated with the high-level theory and basis sets used in the QC calculations for this study.

## 4.1 Energetic ordering

The PIP headgroup structures were ranked according to their energies derived from the geometry optimisation. The energy ranking was done firstly according to each degree of phosphorylation (i.e., mono-, bis-, and tri-phosphorylated PIPs), then within each charge subset. The lowest energy structures were given the ranking 1 and assigned with zero energy. The energy differences to the other structures were calculated with reference to the zero energy structures and ranked accordingly in **Tables 2-4**. The absolute energies of the PIP headgroups can be found in **S.1** of the Supplementary Information.

Table 2: The overall and charge-subset energetic ordering of geometry-optimised phosphatidylinositol monophosphates (PIP).

| PIP | ΔE (kJ $mol^{-1}$) within PIP | Overall energetic ordering | ΔE (kJ $mol^{-1}$) within charge subsets | Energetic ordering | Dipole moment (D) |
|---|---|---|---|---|---|
| 2_PI4P | 0 | 1 | 0 | 1 | 5.21 |
| 2_PI5P | 1.94 | 2 | 1.94 | 2 | 13.64 |
| 2_PI3P | 2.75 | 3 | 2.75 | 3 | 8.89 |
| 3_PI5P | 1257.29 | 4 | 0 | 1 | 19.91 |
| 3_PI3P | 1277.04 | 5 | 19.75 | 2 | 22.10 |
| 3_PI4P | 1282.33 | 6 | 25.05 | 3 | 18.73 |

Table 3: The overall and charge-subset energetic ordering of geometry-optimised phosphatidylinositol bisphosphates ($PIP_2$).

| $PIP_2$ | ΔE (kJ $mol^{-1}$) within $PIP_2$ | Overall energetic ordering | ΔE (kJ $mol^{-1}$) within charge subsets | Energetic ordering | Dipole moment (D) |
|---|---|---|---|---|---|
| 3_PI45P | 0 | 1 | 0 | 1 | 27.86 |
| 3_PI34P | 38.98 | 2 | 38.98 | 2 | 18.43 |
| 3_PI35P | 45.01 | 3 | 45.01 | 3 | 15.08 |
| 4b_PI45P | 1315.81 | 4 | 0 | 1 | 24.66 |
| 4b_PI35P | 1323.00 | 5 | 7.19 | 2 | 33.69 |
| 4b_PI34P | 1336.96 | 6 | 21.15 | 3 | 15.84 |
| 4a_PI45P | 1350.79 | 7 | 34.98 | 4 | 23.01 |
| 4a_PI34P | 1357.32 | 8 | 41.51 | 5 | 41.32 |
| 4a_PI35P | 1373.71 | 9 | 57.90 | 6 | 27.06 |
| 5_PI34P | 2653.12 | 10 | 0 | 1 | 33.15 |
| 5_PI35P | 2658.40 | 11 | 5.28 | 2 | 18.87 |
| 5_PI45P | 2718.08 | 12 | 64.96 | 3 | 46.65 |

Table 4: The overall and charge-subset energetic ordering of geometry-optimised phosphatidylinositol Trisphosphates ($PIP_3$).

| $PIP_3$ | ΔE (kJ $mol^{-1}$) within $PIP_3$ | Overall energetic ordering | ΔE (kJ $mol^{-1}$) within charge subsets | Energetic ordering | Dipole moment (D) |
|---|---|---|---|---|---|
| 4_PI345P | 0 | 1 | 0 | − | 20.82 |
| 5c_PI345P | 1268.85 | 2 | 0 | 1 | |
| 5b_PI345P | 1329.11 | 3 | 60.26 | 2 | 31.59 |
| 5a_PI345P | 1354.71 | 4 | 85.86 | 3 | 31.91 |
| 6b_PI345P | 2625.50 | 5 | 0 | 1 | 25.25 |
| 6c_PI345P | 2715.851 | 6 | 90.341 | 2 | 42.50 |
| 6a_PI345P | 2715.853 | 7 | 90.343 | 3 | 42.51 |
| 7_PI345P | 4012.90 | 8 | 0 | − | 37.57 |

It is found that the energies of the PIP headgroups with the same degree of phosphorylation generally increase with increasing degree of ionisation (net charge), while the relative energies within each charge subset usually vary slightly depending on the position of phosphorylation and the charge distribution.

Larger energy differences are observed within charge subsets with higher net charges. The −2 charge subset of monophosphorylated PIP has a very small energy difference range (3 kJ $mol^{-1}$), while that for the −6 charge subset of trisphosphorylated $PIP_3$ is much greater (90 kJ $mol^{-1}$). Considering a computational error margin of roughly 5 kJ $mol^{-1}$ from the functionals and basis sets used, the energy differences for the higher charged subsets (−4, −5, −6) are more significant.

The energetic ordering of each charge subset can be partly rationalised through analysing the structural properties (**section 4.2**) and charge distributions (**section 4.4**) of the respective PIP headgroup structures.

## 4.2 Structural properties

All 26 optimised PIP headgroup structures are observed to adopt stable chair-like conformations, **Figure 19(a)**, which are consistent with the geometries for PIPs observed in $^{1}$H-NMR spectra and crystal structures.[77] Previous theoretical studies have observed boat-like structures for −6 and −7 charged $PIP_3$ headgroups when their structures were optimised in gas phase with a lower level of theory, **Figure 19(b)**.[74]

(a) (b)

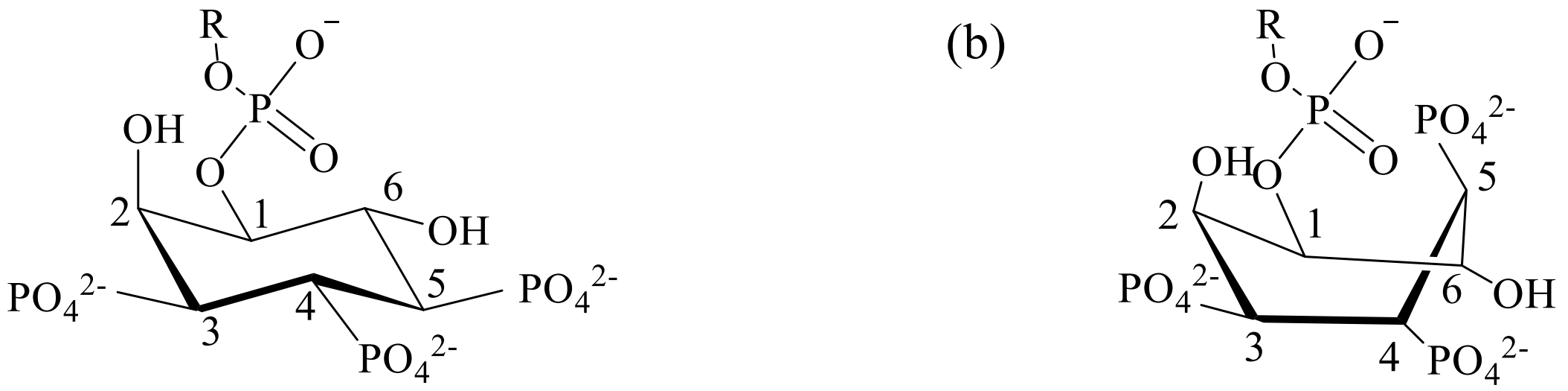

Chair
$7_PI(3,4,5)P_3$

Boat
$7_PI(3,4,5)P_3$

Figure 19: Illustrations of 7_PI345P in its (a) chair conformation and (b) boat conformation.

The stabilisation to the PIP structures afforded by the use of PCM solvent environment in the optimisation calculation is thought to enable this adoption of chair-like structures observed. This is because the solvent stabilised PIP headgroups would be less inclined to twist and form intra-molecular hydrogen bonds (H-bonds) to stabilise their charges compared to when in the gas phase. Nevertheless, intramolecular interactions involving the charged phosphoester groups and the hydroxyl (OH) groups on the inositol rings are still crucial to the structure and stability observed for the optimised PIP headgroups.

The phosphoester substituents on the PIP headgroups can carry either −1 or −2 charges. Stabilising interactions involving these phosphoester groups can be via hydrogen bonding with other phosphates or OH groups on the inositol ring, while destabilising interactions mainly occur though electrostatic charge repulsions. However, it is worth noting that intra-molecular H-bonds formation can also be destabilising if the stability afforded by the former is exceeded by the associated torsional strains or electrostatic repulsion between the highly-charged phosphate groups.

The pair interactions between the phosphate groups on the inositol ring were found by Borkovec *et al* to be negligible in para positions, variable for meta positions, and strongest for ortho positions.[78] The energetic orderings observed for the optimised PIP headgroups (**section 4.1**) illustrate this trend, which could be explained by analysing the stabilising and destabilising interactions present in the structures.

For the monophosphorylated PIPs, the energy differences are small (1−3 kJ mol$^{-1}$) between structures with singly deprotonated $P(OH)O_3^-$ at 3-, 4-, and 5-position. This can be rationalised by the same number of H-bonds (1 H-bond per $P(OH)O_3^-$ group) that are present in each structure, **Figure 20-22(a)**. When the phosphoester substituent is fully deprotonated, the optimised geometries suggest that each $PO_4^{2-}$ requires two H-bonds for stabilisation, which is achieved with its neighbouring OHs on both sides of the ring, **Figure 20-22(b)**. For 3_PI3P, the additional H-bond formation with the axial OH on the 2-position disrupts the latter's ability to form H-bond to stabilise the $P(OMe)O_3^{2-}$ on the 1-postion, **Figure 20(b)**, potentially accounting for the higher energy of the 3_PI3P structure.

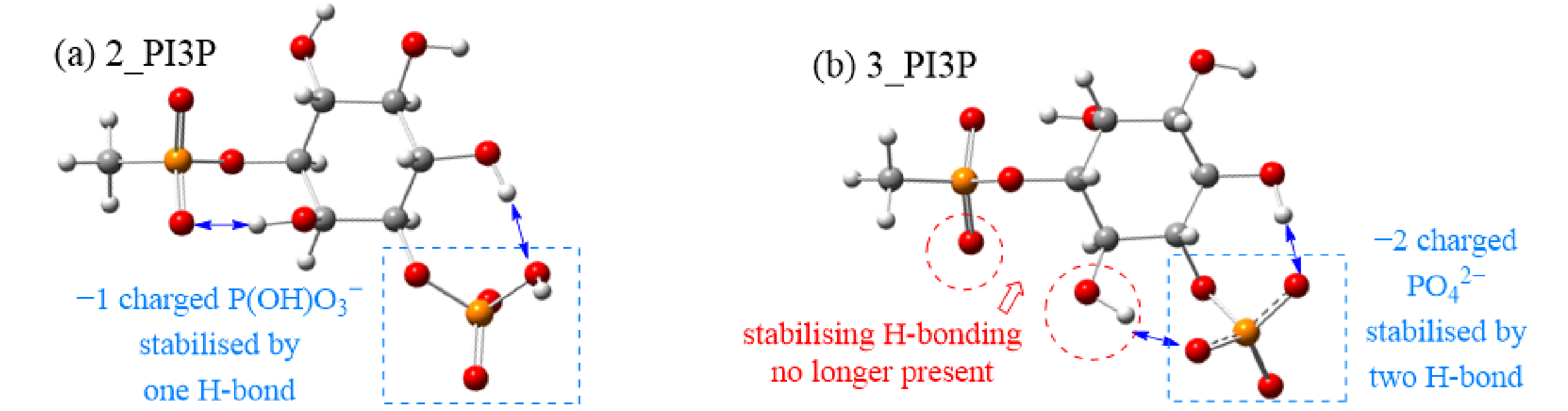

Figure 20: Hydrogen bonding interactions in (a) 2_PI3P and (b) 3_PI3P.

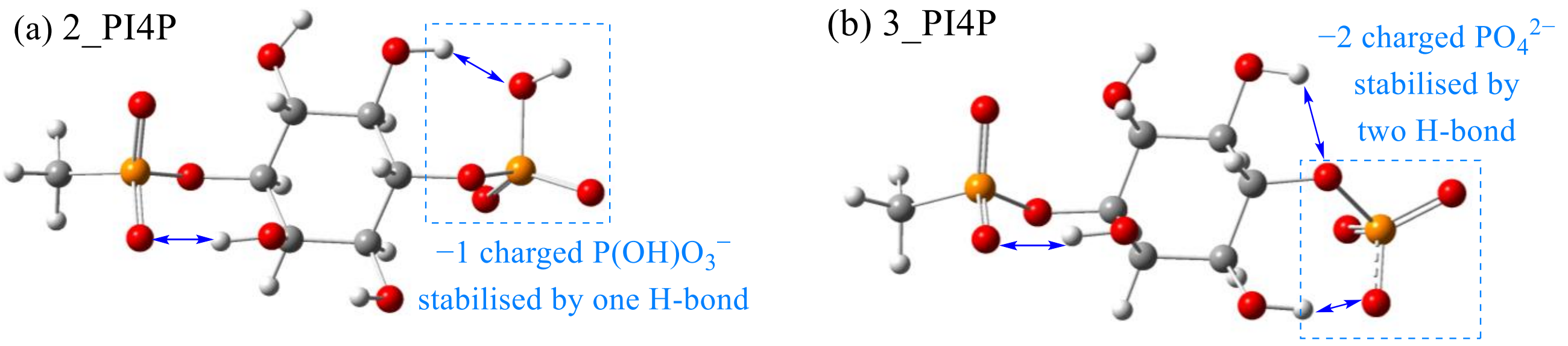

Figure 21: Hydrogen bonding interactions in (a) 2_PI4P and (b) 3_PI4P.

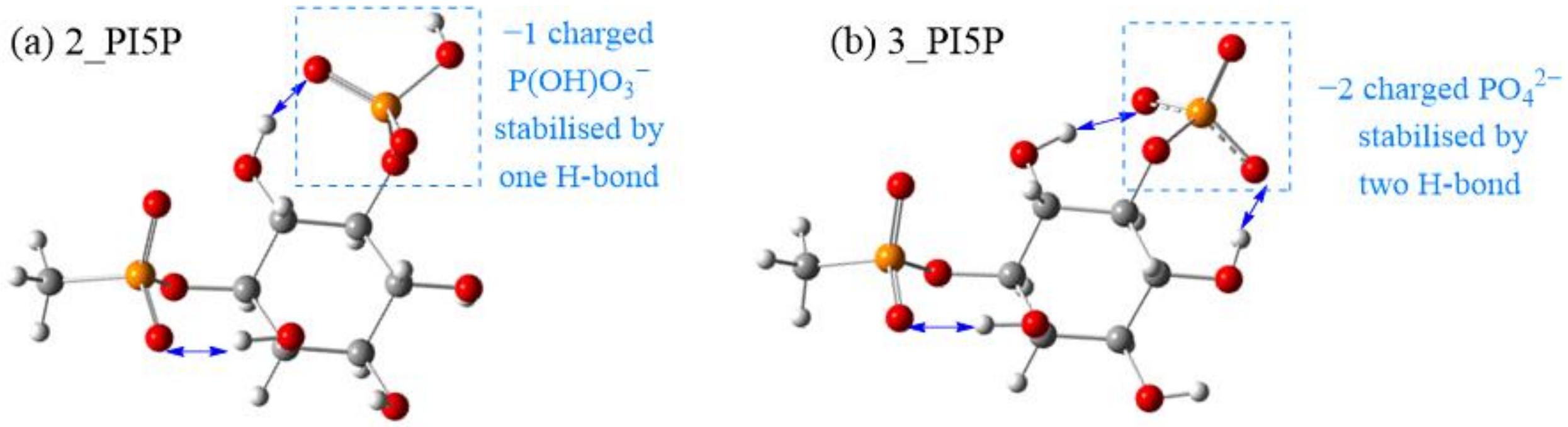

Figure 22: Hydrogen bonding interactions in (a) 2_PI5P and (b) 3_PI5P.

Similarly, the H-bonds between the ortho phosphoester substituents are found to account for the energetic stabilities observed in the bisphosphorylated $PIP_2$. For 3_PI34P and 3_PI45P, the ortho phosphoesters are able to interact strongly and orient themselves to allow for H-bond formation, **Figure 23**. While for 3_PI35P, because the $P(OH)O_3^-$ on the 3- and 5-postions are further apart, they have lower ability to stabilise their charges through forming H-bonds. Instead, it is possible that the two $P(OH)O_3^-$ groups are competing to form H-bond with the same ring OH group on the 4-postion, causing oscillatory behaviour of this OH group that further lowers the stability. This is further supported by *in vitro* studies that suggests weak interactions between phosphoesters in meta positions when separated by an OH group in-between.[78] Moreover, the $P(OH)O_3^-$ on the 5-postion tilts its orientation towards the top face of the ring to interact with the inositol-ring hydrogens, **Figure 24**, creating additional angle strain and steric clash, which may explain the higher energy for 3_PI35P compared to 3_PI34P and 3_PI45P.

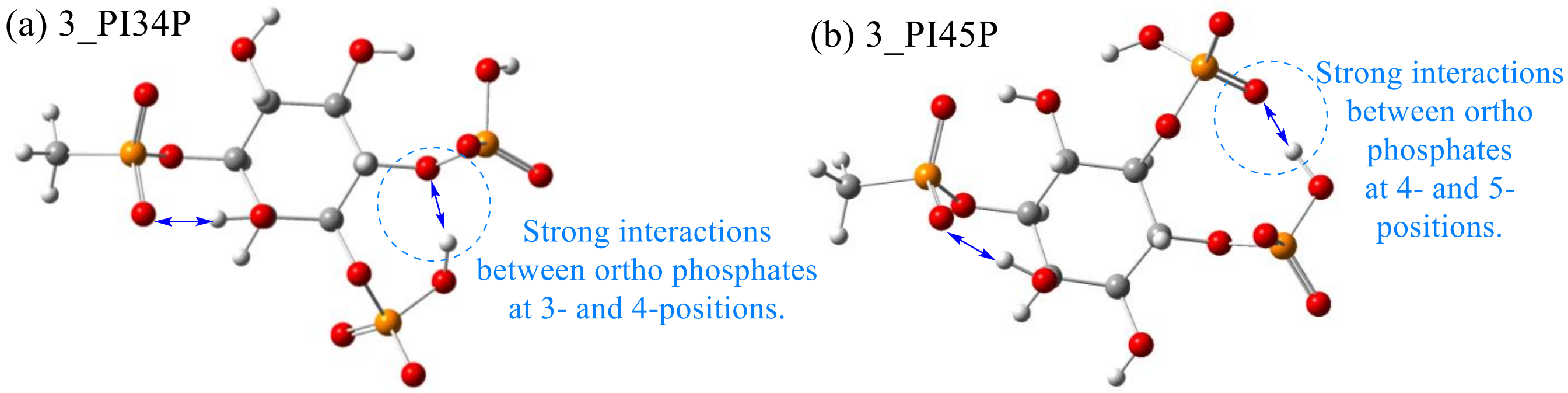

Figure 23: Stabilising hydrogen bonding interactions in (a) 3_PI34P and (b) 3_PI45P.

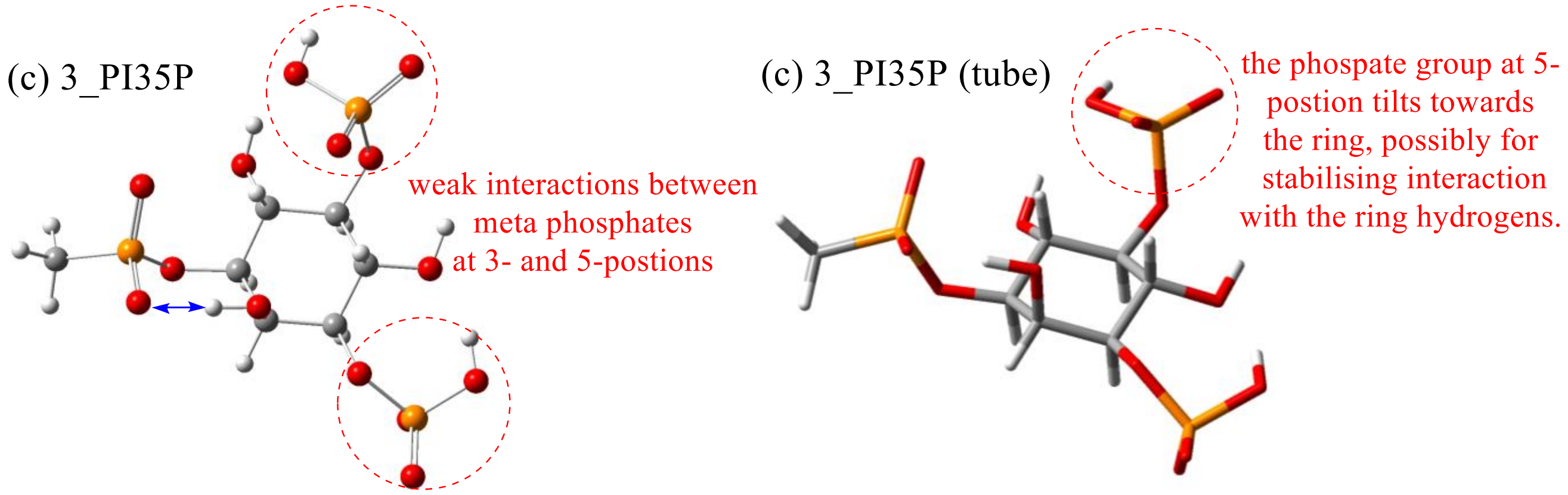

Figure 24: Weak interactions between meta phosphoesters in 3_PI45P.

For the trisphosphorylated $PIP_3$, the energetic ordering can be largely rationalised by the destabilising electrostatic interactions between the multiple doubly deprotonated $PO_4^{2-}$ groups, as well as the stabilising

H-bonds between the singly charged $P(OH)O_3^-$. This is evident in the lowest energy structure 4_PI345P, which is stabilised by a network of intramolecular H-bonds involving the $P(OH)O_3^-$ substituents, **Figure 25(a)**. Whereas the much higher energy of the fully-deprotonated 7_PI345P could be explained by the significant electrostatic repulsion between the three $PO_4^{2-}$ groups and the lack of available proton to facilitate H-bond formation, **Figure 25(b)**.

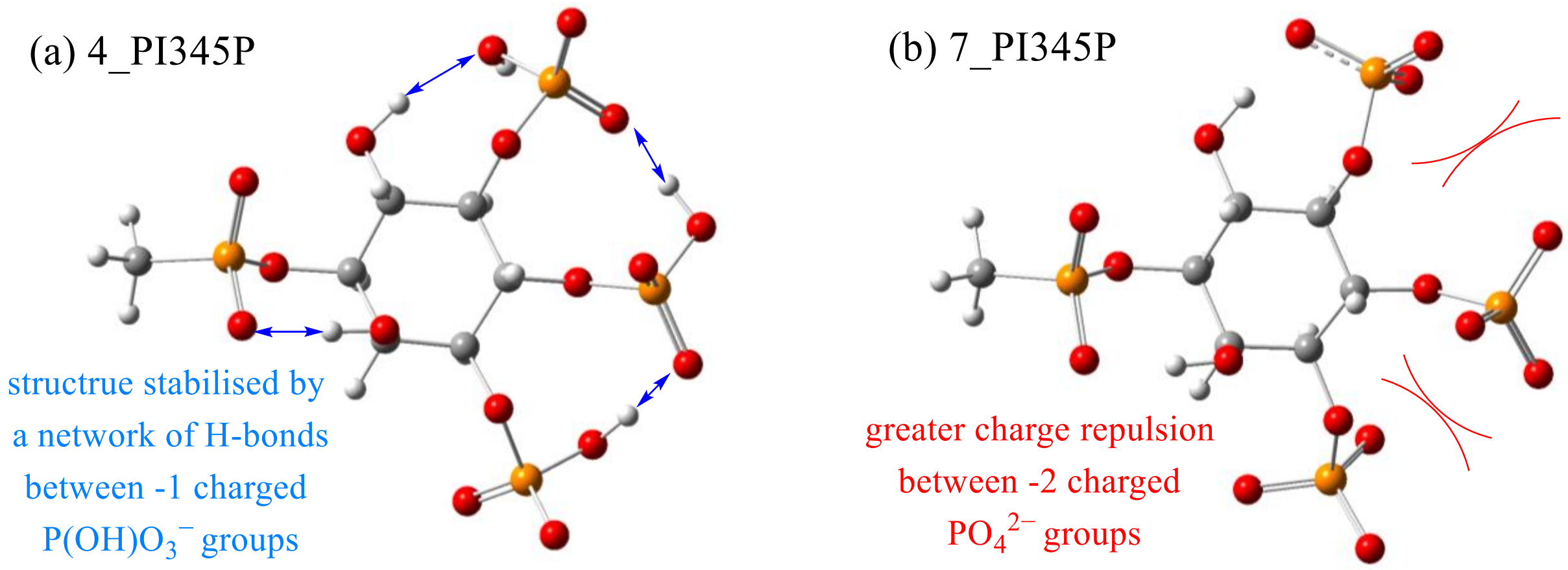

Figure 25: (a) Hydrogen bonding interactions in 4_PI345P and (b) electrostatic repulsion in 7_PI345P.

In the −6 charge subset of $PIP_3$, the most stable structure 6b_PI345P has its two $PO_4^{2-}$ substituents furthest apart (meta). This separation minimise the interactions between the two $PO_4^{2-}$ groups, resulting in its energy being ~90 kJ $mol^{-1}$ lower than both 6a_PI345P and 6c_PI345P. The latter two's structures have ortho $PO_4^{2-}$ groups that repel more strongly. Despite the two $PO_4^{2-}$ substituents are occupying 3,4- and 4,5- positions respectively in 6a_PI345P and 6c_PI345P, these two structures are surprisingly close in energy (differing by 0.02 kJ $mol^{-1}$). The incredibly similar energies suggest that charge repulsion plays a strong role in influencing the energetic stability of these highly-charged $PIP_3$ headgroups.

Despite the apparent correlation between the energetic orderings and the presence of stabilising interactions in the optimised structures, it is worth noting that the ionisation behaviours of PIPs are highly complex, and are often dependent on the exact environmental conditions and the specific biological roles served by the PIPs. Hence, the energetic stability of the respective PIP structures around physiological pH should not be used as the sole guide for determining their charge state probability. Nevertheless, it is encouraging to observe close agreements between the calculated energetic ordering and the experimentally-determined probabilities.

### 4.3 Theory vs experiment

As the optimisations were calculated in implicit water solvent using PCM, the results can be compared with the available experimental data. *In vitro* probabilities for inositol tetrakisphosphates, Ins(1,3,4,5)$P_4$, have been previously established by Borkovec *et al* using $^{31}$P-NMR titration.[78] Ins(1,3,4,5)$P_4$ differs from the PI(3,4,5)$P_3$ headgroups calculated in this work only in the phosphoester substituent at the 1-position of the inositol ring; this phosphoester is not capped in Ins(1,3,4,5)$P_4$, whereas in PI(3,4,5)$P_3$ headgroups it is capped with a terminal methyl group, **Figure 26**. Hence, Ins(1,3,4,5)$P_4$'s experimental probability ordering within each charge subset can be compared with the energetic ordering of the corresponding PI(3,4,5)$P_3$ derived from the calculations, **Table 5**.

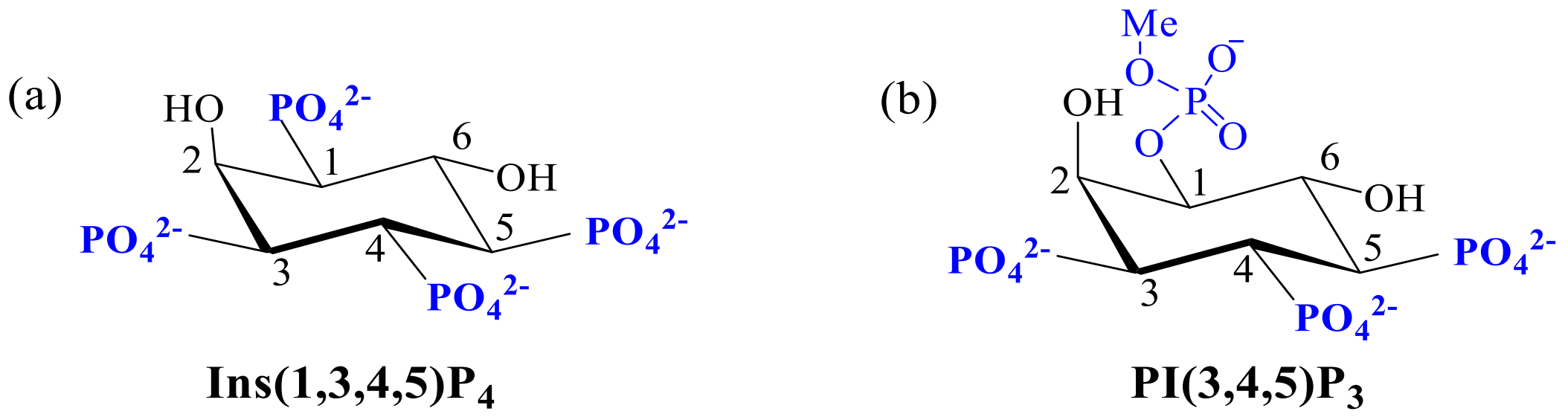

Figure 26: The fully deprotonated structures of (a) Ins(1,3,4,5)$P_4$ and (b) PI(3,4,5)$P_3$.

Table 5: The comparison of the energetic ordering of PI(3,4,5)$P_3$ with the experimental probability of Ins(1,3,4,5)$P_4$.

| $PIP_3$ | ΔE (kJ mol$^{-1}$) within charge subsets | Energetic ordering | Relative *In vitro* probability ordering | *In vitro* probability in % |
|---|---|---|---|---|
| 4_PI345P | 0 | – | – | 100 |
| 5c_PI345P | 0 | 1 | 1 | 35.4 |
| 5b_PI345P | 60.26 | 2 | 2 | 28.8 |
| 5a_PI345P | 85.86 | 3 | 3 | 25.4 |
| | | | All other −5 charged Ins(1,3,4,5)$P_4$ structures probably: | 10.4 |
| 6b_PI345P | 0 | 1 | 1 | 3.8 |
| 6c_PI345P | 90.341 | 2 | 3 | 0.4 |
| 6a_PI345P | 90.343 | 3 | 2 | 0.5 |
| | | | All other −6 charged Ins(1,3,4,5)$P_4$ structures probably: | 95.3 |
| 7_PI345P | 0 | – | – | <0.1 |
| | | | All other −7 charged Ins(1,3,4,5)$P_4$ structures probably: | >99.9 |

The relative energetic ordering from the QC calculations is found to be in good agreement with the corresponding *in vitro* probability ordering. Although a flip in the order is observed for 6a_PI345P and 6c_PI345P, the energy difference calculated (<0.02 kJ $mol^{-1}$) and the *in vitro* probability difference (0.1%) between the reversed pairs are negligible compared to experimental error margins.

Furthermore, the extremely low *in vitro* probability observed for 7_PI345P (<0.1%) is in agreement with the discussion in **Section 4.2**, where the significant electrostatic repulsions between the three charged $PO_4^{2-}$ groups account for the structure's relative high energy.

The QC calculations are found to be able to reproduce similar energetic ordering to the *in vitro* results for the highly-charge $PIP_3$ headgroups at physiological pH. This agreement with experimental data provided confidence in the level of QC theory used in the calculations and the adoption of PCM as the solvation model. Moreover, the results suggested that the energetic properties of the uncapped $PIP_3$ were generally unaffected by the inclusion of the methyl capping group.

## 4.4 ESP and RESP charges

The ESP and RESP charges have been determined for all 26 optimised PIP structures. A full list is provided in **S.2** of Supporting Information.

The charge distributions of the PIP headgroups are visualised by plotting the ESP surfaces. As expected, the negative charges are mainly concentrated on the oxygen atoms in the phosphoester substituents, while the inositol ring carbons carry positive charges. This trend is observed in all ESP surfaces of the optimised PIP structures in **S.1** of Supporting information. A full set of ESP and RESP charges are provided below for 2_PI4P as an example, **Figure 27** & **Table 6**.

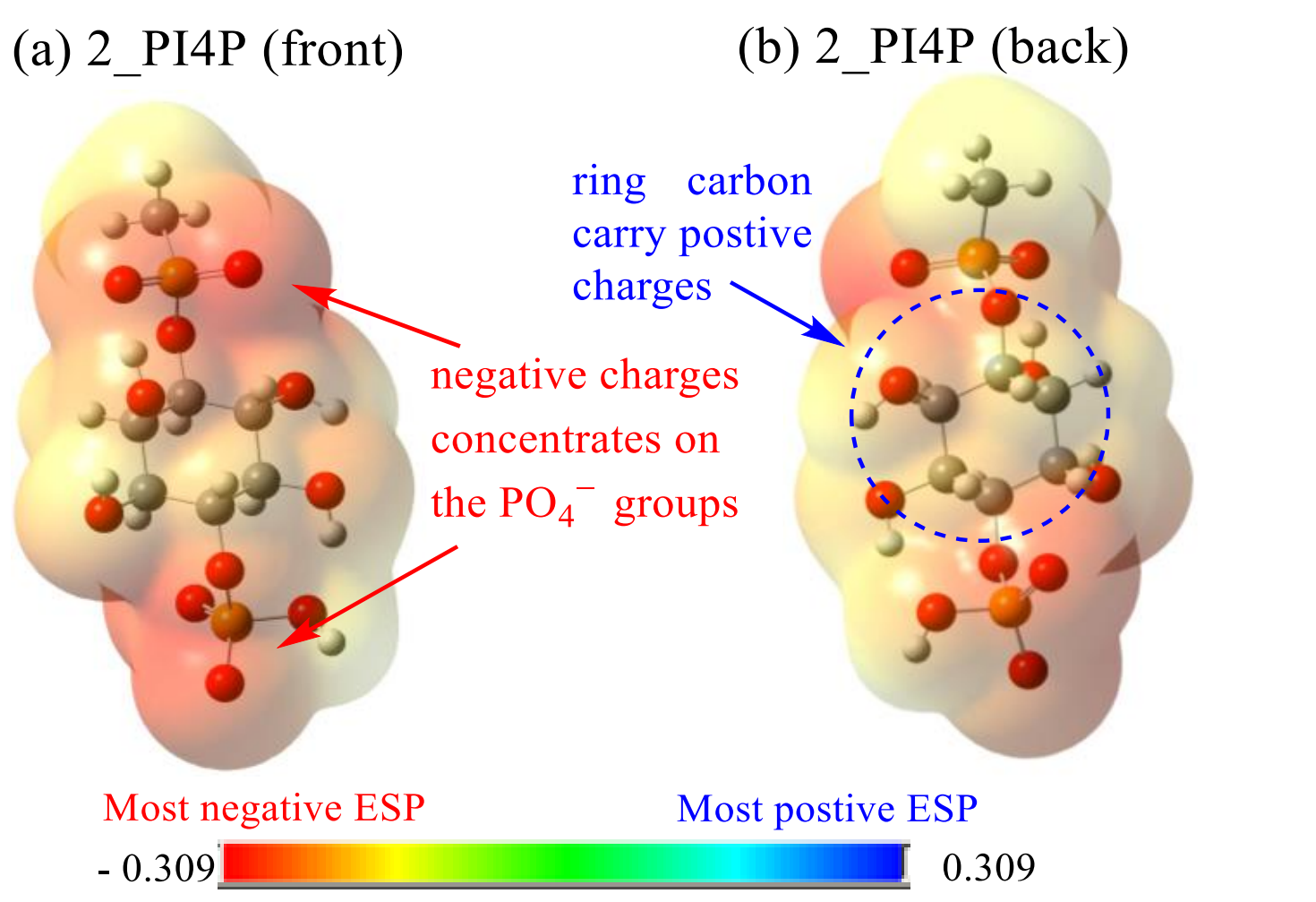

Figure 27 (a-b): ESP surfaces for 2_PI4P (a) front, (b) back.

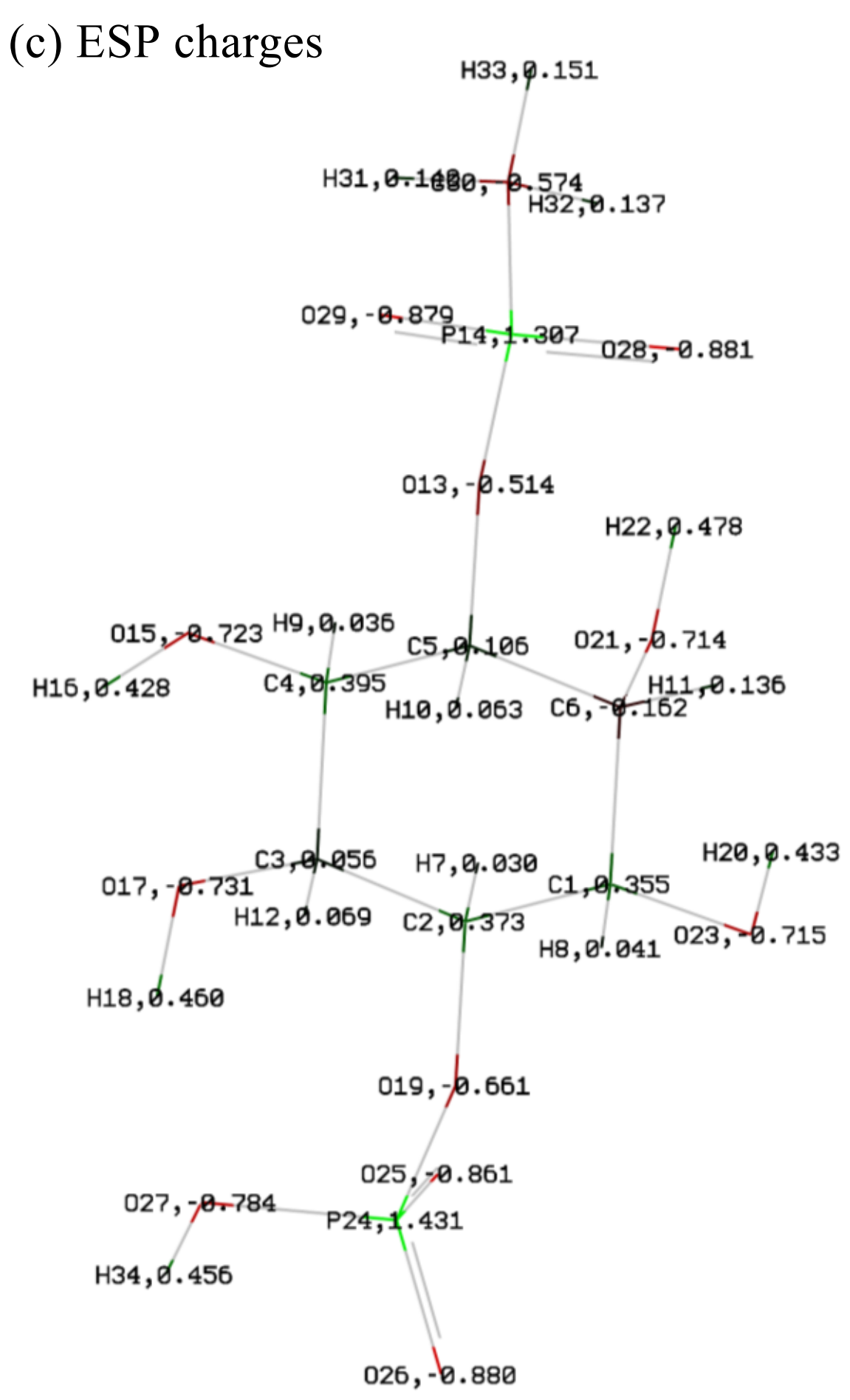

Figure 27 (c): ESP charges for 2_PI4P.

Table 6: RESP charges for 2_PI4P.

| Atom | RESP charge | Atom | RESP charge | Atom | RESP charge |
|---|---|---|---|---|---|
| 1 C1 | 0.069894 | 13 O1 | −0.47937 | 25 O7 | −0.86726 |
| 2 C2 | 0.505434 | 14 P1 | 1.254189 | 26 O8 | −0.86726 |
| 3 C3 | 0.069894 | 15 O2 | −0.63441 | 27 O9 | −0.74524 |
| 4 C4 | 0.162192 | 16 H7 | 0.402027 | 28 O10 | −0.85716 |
| 5 C5 | 0.02026 | 17 O3 | −0.62733 | 29 O11 | −0.85716 |
| 6 C6 | 0.162192 | 18 H8 | 0.370978 | 30 C7 | −0.26653 |
| 7 H1 | −0.00876 | 19 O4 | −0.66682 | 31 H11 | 0.88843 |
| 8 H2 | 0.093455 | 20 H9 | 0.370978 | 32 H12 | 0.88843 |
| 9 H3 | 0.055862 | 21 O5 | −0.63441 | 33 H13 | 0.88843 |
| 10 H4 | 0.079147 | 22 H10 | 0.402027 | 34 H14 | 0.448078 |
| 11 H5 | 0.055862 | 23 O6 | −0.62733 | | |
| 12 H6 | 0.093455 | 24 P2 | 1.395708 | | |

# 5. Conclusions

In this work, every possible ionisation state for the 7 known biologically-active PIPs were investigated computationally to derive charge parameters for use in AMBER. The protocol adopted for the geometry optimisations, headgroup capping, and RESP fitting were consistent with the latest Lipid21 force field.[73] All 26 possible charge permutations for the PIP headgroups have been successfully optimised at the MP2/cc-pVTZ level with the Polarizable Continuum Model. The ESP charges were also calculated at the MP2/cc-pVTZ level and used to derive charge parameters with a two-stage RESP fit.

All optimised geometries adopted stable chair-like structures. The energetic ordering of the optimised structures were rationalised through analysis of the optimised structures and identifying key intramolecular interactions. Intramolecular hydrogen bonding and electrostatic repulsions were found to be the dominant forces in determining the structural features and stabilities of the PIP headgroups. The theoretical energetic ordering obtained were also found to be in close agreement with the available *in vitro* data,[78] providing validation to the level of theory and solvation model used in the calculations.

The optimisation and charge derivation conducted in this work represent the first full set of charged parameters for PIPs obtained at the high-level of theory stipulated in the latest Lipid21 force field. This comprehensive library of charges for PIP headgroups will serve as an extension to the Lipid 21 force field,[73] which will provide a basis for studying PIP-containing biological systems using MD simulations with AMBER.

## 6. Future Work

The most immediate work is to further refine the RESP charges by averaging multiple individual fits to produce Boltzmann-averaged charges over a conformational ensemble. The subsequent assignments of additional parameters (bond, angle, torsion and VDW) are expected to be relatively straightforward using the pre-defined atom types in Lipid21.[73] The quality of the parameters are then to be evaluated through MD simulations of benchmark systems with AMBER, where calculated results will be compared with available experimental data.

After validation, the PIP parameters can be formally incorporated into the Lipid21 force field and used to investigate a broad range of biological systems involving PIPs. For instance, the role PIPs play in regulating membrane trafficking[11] or facilitating intracellular signalling pathways[12] could be explored with MD simulations of the relevant membrane components. In addition, MD studies could be used to obtain valuable atomic level details for important PIP-protein interactions that have yet to be explored or are difficult to study experimentally.[13]

In particular, it would be of great interest to revisit the MD simulations conducted in the Gould group on PI(3,4,5)$P_3$'s interactions with the protein kinase B (PKB) domain, where the generic PIP parameters used were unable to reproduce many critical experimental results.[74]

The specificity and affinity between PKB domains and specific PIP species are of significant research importance due to the association of PKB activations with multiple cancer types.[9] However, experimental investigations of these systems are difficult and expensive,[79] resulting in very few *in vitro* results available in the literature.[80] Hence, MD studies of these systems can be invaluable in providing crucial information about the PKB PH domains and their interactions with PIP lipids. Future work can thus utilise the PIP parameters developed in this work and conduct MD studies to unveil insights on the crucial interactions present between PKB and PI(3,4,5)$P_3$.

## 7. Acknowledgement

I would like to sincerely thank Prof Ian Gould for the opportunity to work in his lab, as well as his valuable guidance and feedback throughout the project. In addition, I would like to thank my group members Tony Yang, Charlie Brown, and Sam Gubbins for the fruitful discussions and support.

## References

1. Berg, J. M.; Tymoczko, J. L.; Stryer, L. Biochemistry. **2002**, 373-379.

2. Lombard, J. Once upon a time the cell membranes: 175 years of cell boundary research. *Biology direct* **2014**, *9*, 1-35.

3. Leonard, A. N.; Wang, E.; Monje-Galvan, V.; Klauda, J. B. Developing and testing of lipid force fields with applications to modeling cellular membranes. *Chem. Rev.* **2019**, *119*, 6227-6269.

4. Van Meer, G.; Voelker, D. R.; Feigenson, G. W. Membrane lipids: where they are and how they behave. *Nature reviews Molecular cell biology* **2008**, *9*, 112-124.

5. Tanford, C. *The hydrophobic effect: formation of micelles and biological membranes 2d ed;* J. Wiley.: 1980.

6. Cevc, G.; Marsh, D. *Phospholipid bilayers: physical principles and models;* Wiley: 1987; .

7. van Meer, G.; Voelker, D. R.; Feigenson, G. W. Membrane lipids: where they are and how they behave. *Nat Rev Mol Cell Biol* **2008**, *9*, 112-124.

8. Lupyan, D.; Mezei, M.; Logothetis, D. E.; Osman, R. A molecular dynamics investigation of lipid bilayer perturbation by PIP2. *Biophys. J.* **2010**, *98*, 240-247.

9. Lemmon, M. A. Membrane recognition by phospholipid-binding domains. *Nature reviews Molecular cell biology* **2008**, *9*, 99-111.

10. Rusten, T. E.; Stenmark, H. Analyzing phosphoinositides and their interacting proteins. *Nature methods* **2006**, *3*, 251-258.

11. Gillooly, D. J.; Simonsen, A.; Stenmark, H. Cellular functions of phosphatidylinositol 3-phosphate and FYVE domain proteins. *Biochem J* **2001**, *355*, 249-258.

12. Falkenburger, B. H.; Jensen, J. B.; Dickson, E. J.; Suh, B.; Hille, B. Phosphoinositides: lipid regulators of membrane proteins. *J Physiol* **2010**, *588*, 3179-3185.

13. Ma, Q.; Zhu, C.; Zhang, W.; Ta, N.; Zhang, R.; Liu, L.; Feng, D.; Cheng, H.; Liu, J.; Chen, Q. Mitochondrial PIP3-binding protein FUNDC2 supports platelet survival via AKT signaling pathway. *Cell Death Differ* **2019**, *26*, 321-331.

14. Pendaries, C.; Tronchère, H.; Plantavid, M.; Payrastre, B. Phosphoinositide signaling disorders in human diseases. *FEBS Lett.* **2003**, *546*, 25-31.

15. Rosen, S. A.; Gaffney, P. R.; Gould, I. R. A theoretical investigation of inositol 1, 3, 4, 5-tetrakisphosphate. *Physical Chemistry Chemical Physics* **2011**, *13*, 1070-1081.

16. Borkovec, M.; Spiess, B. Microscopic ionization mechanism of inositol tetrakisphosphates. *Physical Chemistry Chemical Physics* **2004**, *6*, 1144-1151.

17. Bernardi, R. C.; Melo, M. C. R.; Schulten, K. Biochimica et biophysica acta. *Biochimica et biophysica acta* **1961**, *1850*, 872-877.

18. Hunter, M. S.; DePonte, D. P.; Shapiro, D. A.; Kirian, R. A.; Wang, X.; Starodub, D.; Marchesini, S.; Weierstall, U.; Doak, R. B.; Spence, J. C. H.; Fromme, P. X-ray Diffraction from Membrane Protein Nanocrystals. *Biophysical journal* **2011**, *100*, 198-206.

19. Levental, I.; Cebers, A.; Janmey, P. A. Combined electrostatics and hydrogen bonding determine intermolecular interactions between polyphosphoinositides. *J. Am. Chem. Soc.* **2008**, *130*, 9025-9030.

20. Rosenhouse-Dantsker, A.; Logothetis, D. E. Molecular characteristics of phosphoinositide binding. *Pflügers Archiv-European Journal of Physiology* **2007**, *455*, 45-53.

21. Sauer, K.; Huang, Y. H.; Lin, H.; Sandberg, M.; Mayr, G. W. Phosphoinositide and Inositol Phosphate Analysis in Lymphocyte Activation. *Curr Protoc Immunol* **2009**, *0 11*, Unit11.1.

22. Furse, S.; Mak, L.; Tate, E. W.; Templer, R. H.; Ces, O.; Woscholski, R.; Gaffney, P. R. Synthesis of unsaturated phosphatidylinositol 4-phosphates and the effects of substrate unsaturation on Sop B phosphatase activity. *Organic & biomolecular chemistry* **2015**, *13*, 2001-2011.

23. M. Joffrin, A.; M. Saunders, A.; Barneda, D.; Flemington, V.; L. Thompson, A.; J. Sanganee, H.; J. Conway, S. Development of isotope-enriched phosphatidylinositol-4- and 5-phosphate cellular mass spectrometry probes. *Chemical Science* **2021**, *12*, 2549-2557.

24. Marrink, S. J.; De Vries, A. H.; Tieleman, D. P. Lipids on the move: simulations of membrane pores, domains, stalks and curves. *Biochimica et Biophysica Acta (BBA)-Biomembranes* **2009**, *1788*, 149-168.

25. Marrink, S. J.; Corradi, V.; Souza, P. C.; Ingólfsson, H. I.; Tieleman, D. P.; Sansom, M. S. Computational modeling of realistic cell membranes. *Chem. Rev.* **2019**, *119*, 6184-6226.

26. Izgorodina, E. I.; Seeger, Z. L.; Scarborough, D. L.; Tan, S. Y. Quantum chemical methods for the prediction of energetic, physical, and spectroscopic properties of ionic liquids. *Chem. Rev.* **2017**, *117*, 6696-6754.

27. Friesner, R. A.; Guallar, V. Ab initio quantum chemical and mixed quantum mechanics/molecular mechanics (QM/MM) methods for studying enzymatic catalysis. *Annu.Rev.Phys.Chem.* **2005**, *56*, 389-427.

28. Izgorodina, E. I.; Seeger, Z. L.; Scarborough, D. L. A.; Tan, S. Y. S. Quantum Chemical Methods for the Prediction of Energetic, Physical, and Spectroscopic Properties of Ionic Liquids. *Chem. Rev.* **2017**, *117*, 6696-6754.

29. Laaksonen, L.; Pyykkö, P.; Sundholm, D. Fully numerical Hartree-Fock methods for molecules. *Computer physics reports* **1986**, *4*, 313-344.

30. Della Sala, F.; Görling, A. Efficient localized Hartree–Fock methods as effective exact-exchange Kohn–Sham methods for molecules. *J. Chem. Phys.* **2001**, *115*, 5718-5732.

31. Møller, C.; Plesset, M. S. Note on an approximation treatment for many-electron systems. *Physical review* **1934**, *46*, 618.

32. Head-Gordon M, Pople J A, Frisch M J. MP2 energy evaluation by direct methods. *Chemical Physics Letters* **1988**, *153*, 503-506.

33. Obot, I. B.; Macdonald, D. D.; Gasem, Z. M. Density functional theory (DFT) as a powerful tool for designing new organic corrosion inhibitors. Part 1: an overview. *Corros. Sci.* **2015**, *99*, 1-30.

34. Werner, H.; Manby, F. R.; Knowles, P. J. Fast linear scaling second-order Møller-Plesset perturbation theory (MP2) using local and density fitting approximations. *J. Chem. Phys.* **2003**, *118*, 8149-8160.

35. Rassolov, V. A.; Pople, J. A.; Ratner, M. A.; Windus, T. L. 6-31G* basis set for atoms K through Zn. *J. Chem. Phys.* **1998**, *109*, 1223-1229.

36. Alkorta, I.; Elguero, J. Review on DFT and ab initio calculations of scalar coupling constants. *International Journal of Molecular Sciences* **2003**, *4*, 64-92.

37. Huang, J.; Kertesz, M. Intermolecular transfer integrals for organic molecular materials: can basis set convergence be achieved? *Chemical physics letters* **2004**, *390*, 110-115.

38. Gabl, S.; Schröder, C.; Steinhauser, O. Computational studies of ionic liquids: Size does matter and time too. *J. Chem. Phys.* **2012**, *137*, 094501.

39. Peterson, K. A.; Dunning, T. H. Accurate correlation consistent basis sets for molecular core–valence correlation effects: The second row atoms Al–Ar, and the first row atoms B–Ne revisited. *J. Chem. Phys.* **2002**, *117*, 10548-10560.

40. Peterson, K. A.; Figgen, D.; Goll, E.; Stoll, H.; Dolg, M. Systematically convergent basis sets with relativistic pseudopotentials. II. Small-core pseudopotentials and correlation consistent basis sets for the post-d group 16–18 elements. *J. Chem. Phys.* **2003**, *119*, 11113-11123.

41. Fehér, P. P.; Stirling, A. Assessment of reactivities with explicit and implicit solvent models: QM/MM and gas-phase evaluation of three different Ag-catalysed furan ring formation routes. *New J. Chem.* **2019**, *43*, 15706-15713.

42. Kamerlin, S. C. L.; Haranczyk, M.; Warshel, A. Are Mixed Explicit/Implicit Solvation Models Reliable for Studying Phosphate Hydrolysis? A Comparative Study of Continuum, Explicit and Mixed Solvation Models. *ChemPhysChem* **2009**, *10*, 1125-1134.

43. Cramer, C. J.; Truhlar, D. G. Implicit solvation models: equilibria, structure, spectra, and dynamics. *Chem. Rev.* **1999**, *99*, 2161-2200.

44. Marenich, A. V.; Cramer, C. J.; Truhlar, D. G. Universal Solvation Model Based on Solute Electron Density and on a Continuum Model of the Solvent Defined by the Bulk Dielectric Constant and Atomic Surface Tensions. *J. Phys. Chem. B* **2009**, *113*, 6378-6396.

45. Mennucci, B. Polarizable continuum model. *WIREs Computational Molecular Science* **2012**, *2*, 386-404.

46. Bernales, V. S.; Marenich, A. V.; Contreras, R.; Cramer, C. J.; Truhlar, D. G. Quantum mechanical continuum solvation models for ionic liquids. *The Journal of Physical Chemistry B* **2012**, *116*, 9122-9129.

47. Hunt, P. A. Quantum Chemical Modeling of Hydrogen Bonding in Ionic Liquids. *Top. Curr. Chem.* **2017**, *375*, 59.

48. Vanommeslaeghe, K.; Guvench, O. Molecular mechanics. *Curr. Pharm. Des.* **2014**, *20*, 3281-3292.

49. Senn, H. M.; Thiel, W. QM/MM methods for biological systems. *Atomistic approaches in modern biology* **2006**, 173-290.

50. Wereszczynski, J.; McCammon, J. A. Statistical mechanics and molecular dynamics in evaluating thermodynamic properties of biomolecular recognition. *Q. Rev. Biophys.* **2012**, *45*, 1.

51. Stillinger, F. H.; Rahman, A. Improved simulation of liquid water by molecular dynamics. *J. Chem. Phys.* **1974**, *60*, 1545-1557.

52. Spreiter, Q.; Walter, M. Classical molecular dynamics simulation with the Velocity Verlet algorithm at strong external magnetic fields. *Journal of Computational Physics* **1999**, *152*, 102-119.

53. Rapaport, D. C. *The art of molecular dynamics simulation;* Cambridge university press: 2004; .

54. Hospital, A.; Goñi, J. R.; Orozco, M.; Gelpí, J. L. Molecular dynamics simulations: advances and applications. *Adv Appl Bioinform Chem* **2015**, *8*, 37-47.

55. Alder, B. J.; Wainwright, T. E. Studies in molecular dynamics. I. General method. *J. Chem. Phys.* **1959**, *31*, 459-466.

56. Chandler D. Introduction to modern statistical Mechanics. *Oxford University Press, Oxford, UK*, 1987, 40.

57. Wood, W. W.; Erpenbeck, J. J.; Baker Jr, G. A.; Johnson, J. D. Molecular dynamics ensemble, equation of state, and ergodicity. *Physical Review E* **2000**, *63*, 011106.

58. Patrascioiu, A. The ergodic hypothesis. *Proceedings of Los Almos Science Special issue* **1987**, 263-279.

59. Oostenbrink, C.; Villa, A.; Mark, A. E.; Van Gunsteren, W. F. A biomolecular force field based on the free enthalpy of hydration and solvation: the GROMOS force-field parameter sets 53A5 and 53A6. *Journal of computational chemistry* **2004**, *25*, 1656-1676.

60. MacKerell Jr, A. D.; Bashford, D.; Bellott, M.; Dunbrack Jr, R. L.; Evanseck, J. D.; Field, M. J.; Fischer, S.; Gao, J.; Guo, H.; Ha, S. All-atom empirical potential for molecular modeling and dynamics studies of proteins. *The journal of physical chemistry B* **1998**, *102*, 3586-3616.

61. Salomon-Ferrer, R.; Case, D. A.; Walker, R. C. An overview of the Amber biomolecular simulation package. *Wiley Interdisciplinary Reviews: Computational Molecular Science* **2013**, *3*, 198-210.

62. Pohle, W.; Gauger, D. R.; Bohl, M.; Mrazkova, E.; Hobza, P. Lipid hydration: headgroup CH moieties are involved in water binding. *Biopolymers: Original Research on Biomolecules* **2004**, *74*, 27-31.

63. Venable, R. M.; Luo, Y.; Gawrisch, K.; Roux, B.; Pastor, R. W. Simulations of anionic lipid membranes: development of interaction-specific ion parameters and validation using NMR data. *The journal of physical chemistry B* **2013**, *117*, 10183-10192.

64. Wang, J.; Wolf, R. M.; Caldwell, J. W.; Kollman, P. A.; Case, D. A. Development and testing of a general amber force field. *Journal of Computational Chemistry* **2004**, *25*, 1157-1174.

65. Cornell, W. D.; Cieplak, P.; Bayly, C. I.; Gould, I. R.; Merz, K. M.; Ferguson, D. M.; Spellmeyer, D. C.; Fox, T.; Caldwell, J. W.; Kollman, P. A. A second generation force field for the simulation of proteins, nucleic acids, and organic molecules. *J. Am. Chem. Soc.* **1995**, *117*, 5179-5197.

66. Hornak, V.; Abel, R.; Okur, A.; Strockbine, B.; Roitberg, A.; Simmerling, C. Comparison of multiple Amber force fields and development of improved protein backbone parameters. *Proteins: Structure, Function, and Bioinformatics* **2006**, *65*, 712-725.

67. Kirschner, K. N.; Woods, R. J. Solvent interactions determine carbohydrate conformation. *Proceedings of the National Academy of Sciences* **2001**, *98*, 10541-10545.

68. Woods, R. J.; Dwek, R. A.; Edge, C. J.; Fraser-Reid, B. Molecular mechanical and molecular dynamic simulations of glycoproteins and oligosaccharides. 1. GLYCAM_93 parameter development. *J. Phys. Chem.* **1995**, *99*, 3832-3846.

69. Dickson, C. J.; Madej, B. D.; Skjevik, Å A.; Betz, R. M.; Teigen, K.; Gould, I. R.; Walker, R. C. Lipid14: the amber lipid force field. *Journal of chemical theory and computation* **2014**, *10*, 865-879.

70. Wang, J.; Wolf, R. M.; Caldwell, J. W.; Kollman, P. A.; Case, D. A. Development and testing of a general amber force field. *Journal of computational chemistry* **2004**, *25*, 1157-1174.

71. Ollila, O. S.; Pabst, G. Atomistic resolution structure and dynamics of lipid bilayers in simulations and experiments. *Biochimica et Biophysica Acta (BBA)-Biomembranes* **2016**, *1858*, 2512-2528.

72. Botan, A.; Favela-Rosales, F.; Fuchs, P. F.; Javanainen, M.; Kanduč, M.; Kulig, W.; Lamberg, A.; Loison, C.; Lyubartsev, A.; Miettinen, M. S. Toward atomistic resolution structure of phosphatidylcholine headgroup and glycerol backbone at different ambient conditions. *The Journal of Physical Chemistry B* **2015**, *119*, 15075-15088.

73. Dickson, C.; Walker, R.; Gould, I. Lipid21: Complex Lipid Membrane Simulations with AMBER.

74. Rosen, S. Investigating the interaction of the PKB PH domain with inositol phosphatebased compounds, Imperial College London, 2013.

75. Gaussian 16, Revision C.01, M. J. Frisch, G. W. Trucks, H. B. Schlegel, G. E. Scuseria, M. A. Robb, J. R. Cheeseman, G. Scalmani, V. Barone, G. A. Petersson, H. Nakatsuji, X. Li, M. Caricato, A. V. Marenich, J. Bloino, B. G. Janesko, R. Gomperts, B. Mennucci, H. P. Hratchian, J. V. Ortiz, A. F. Izmaylov, J. L. Sonnenberg, D. Williams-Young, F. Ding, F. Lipparini, F. Egidi, J. Goings, B. Peng, A. Petrone, T. Henderson, D. Ranasinghe, V. G. Zakrzewski, J. Gao, N. Rega, G. Zheng, W. Liang, M. Hada, M. Ehara, K. Toyota, R. Fukuda, J. Hasegawa, M. Ishida, T. Nakajima, Y. Honda, O. Kitao, H. Nakai, T. Vreven, K. Throssell, J. A. Montgomery, Jr., J. E. Peralta, F. Ogliaro, M. J. Bearpark, J. J. Heyd, E. N. Brothers, K. N. Kudin, V. N. Staroverov, T. A. Keith, R. Kobayashi, J. Normand, K. Raghavachari, A. P. Rendell, J. C. Burant, S. S. Iyengar, J. Tomasi, M. Cossi, J. M. Millam, M. Klene, C. Adamo, R. Cammi, J. W. Ochterski, R. L. Martin, K. Morokuma, O. Farkas, J. B. Foresman, and D. J. Fox, Gaussian, Inc., Wallingford CT, 2019.

76. GaussView, Version 6.0.16, Dennington, Roy; Keith, Todd A.; Millam, John M. Semichem Inc., Shawnee Mission, KS, 2016.

77. Milburn, C. C.; Deak, M.; Kelly, S. M.; Price, N. C.; Alessi, D. R.; Van Aalten, D. M. Binding of phosphatidylinositol 3, 4, 5-trisphosphate to the pleckstrin homology domain of protein kinase B induces a conformational change. *Biochem. J.* **2003**, *375*, 531-538.

78. Borkovec, M.; Spiess, B. Microscopic ionization mechanism of inositol tetrakisphosphates. *Physical Chemistry Chemical Physics* **2004**, *6*, 1144-1151.

79. Deprez, J.; Vertommen, D.; Alessi, D. R.; Hue, L.; Rider, M. H. Phosphorylation and activation of heart 6-phosphofructo-2-kinase by protein kinase B and other protein kinases of the insulin signaling cascades. *J. Biol. Chem.* **1997**, *272*, 17269-17275.

80. Conway, S. J.; Miller, G. J. Biology-enabling inositol phosphates, phosphatidylinositol phosphates and derivatives. *Nat. Prod. Rep.* **2007**, *24*, 687-707.

# Supporting Information

## S.1 Summary of optimisation and ESP jobs

| Molecule | Absolute Energy (Hartree) | diople moment (DB) | ESP (front) | ESP (back) |
|---|---|---|---|---|
| 2_PI3P | −1783.264991 | 8.887721 | 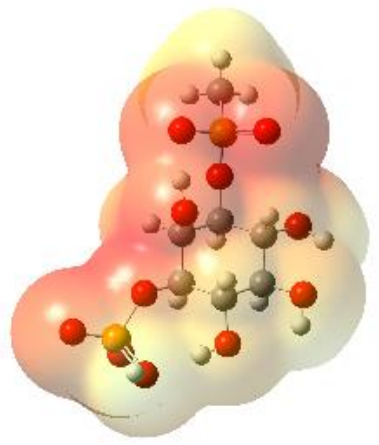 | 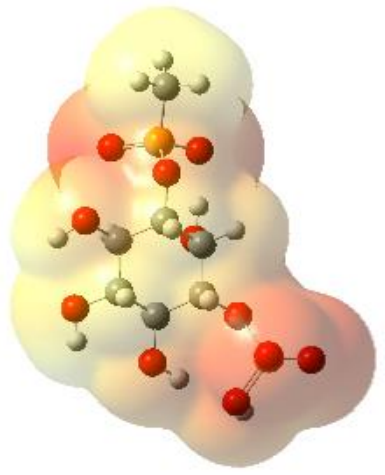 |
| 2_PI4P | −1783.26604 | 5.20524 | 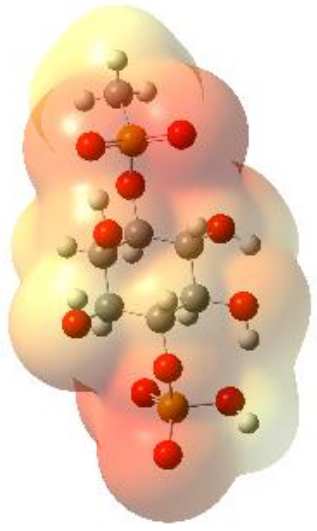 | 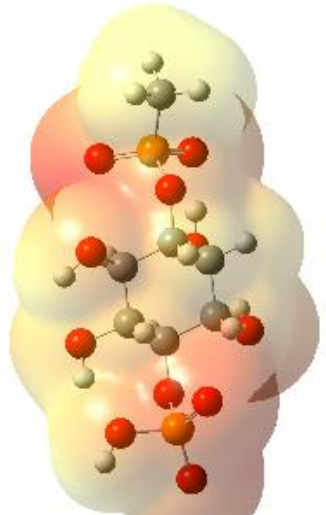 |
| 2_PI5P | −1783.265302 | 13.640085 | 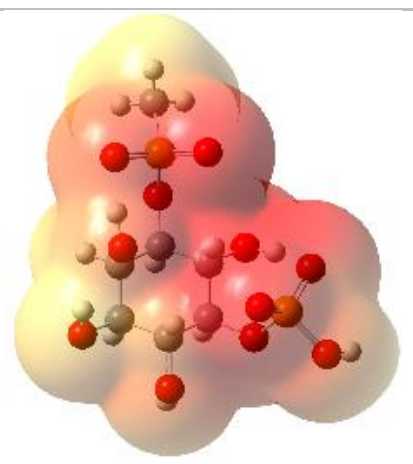 | 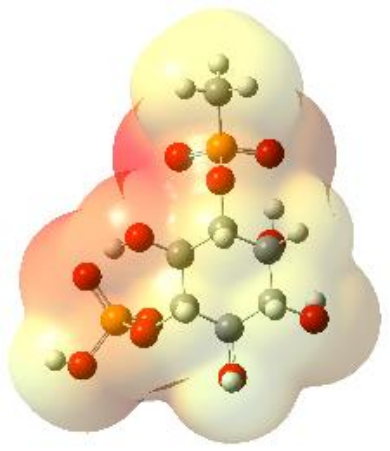 |
| 3_PI3P | −1782.779642 | 22.097943 | 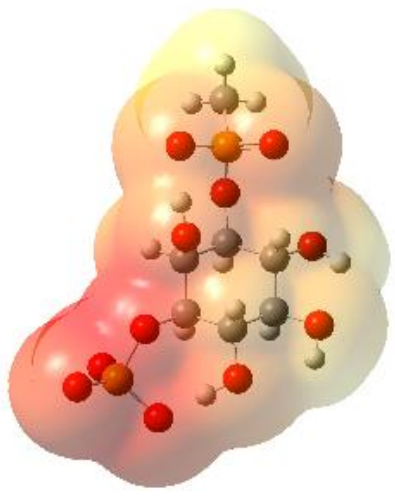 | 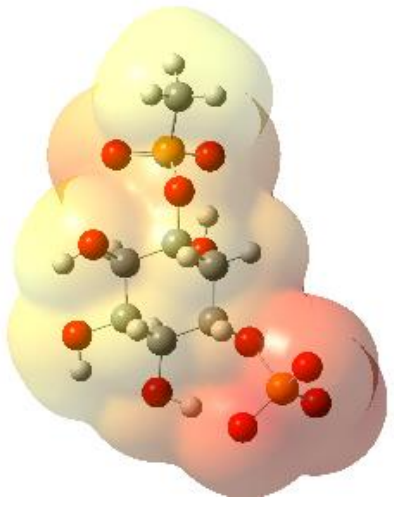 |
| 3_PI4P | −1782.777625 | 18.728722 | 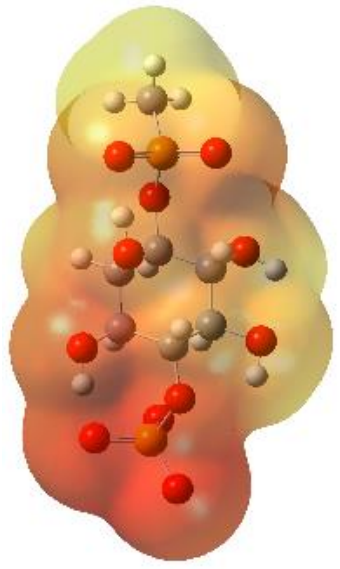 | 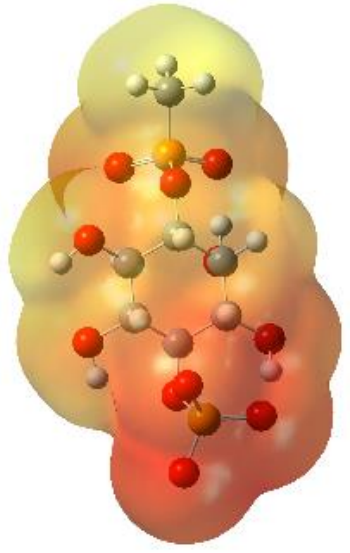 |
| 3_PI5P | −1782.787165 | 19.910688 | 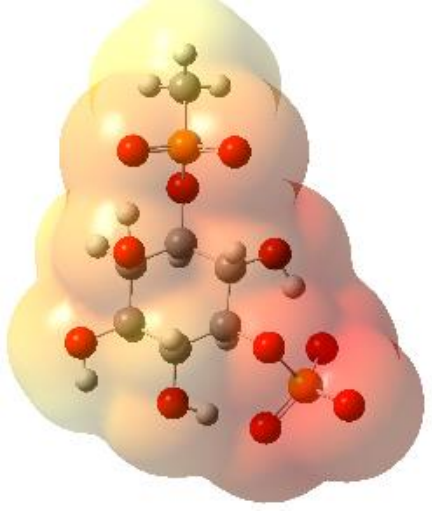 | 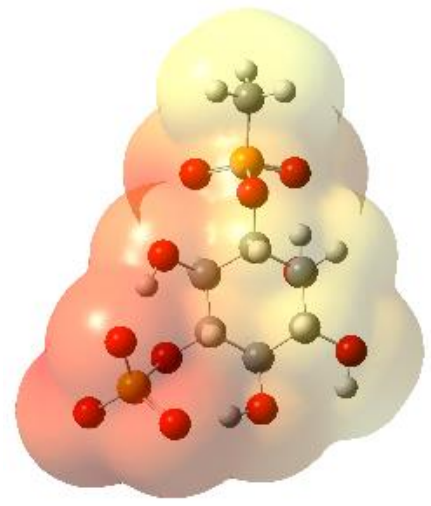 |
| 3_PI34P | −2349.828541 | 18.433225 | 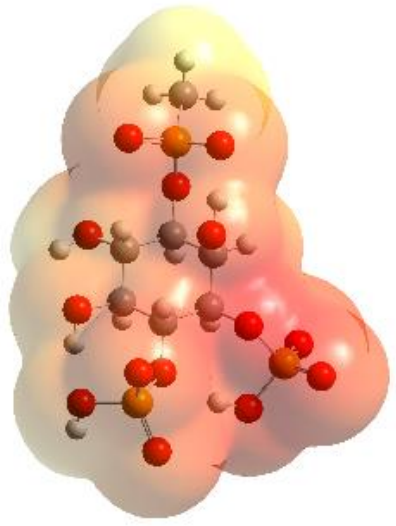 | 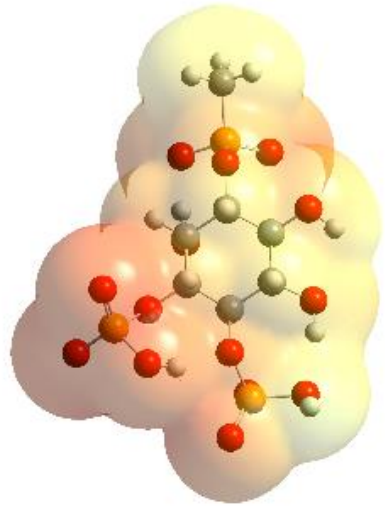 |
| 3_PI35P | −2349.826242 | 15.081088 | 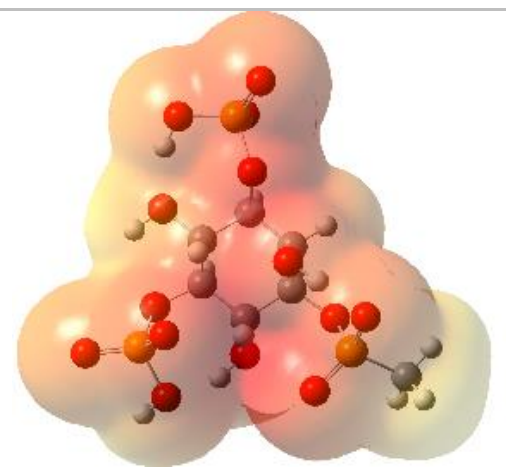 | 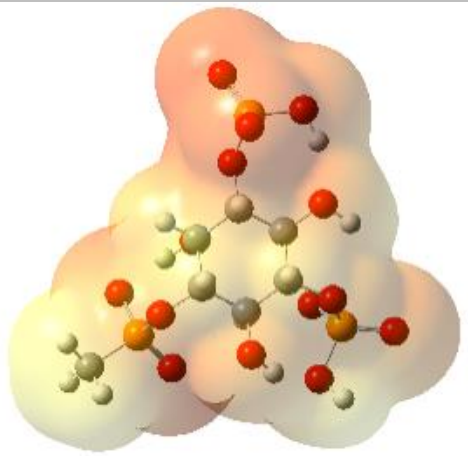 |
| 3_PI45P | −2349.843386 | 27.858176 | 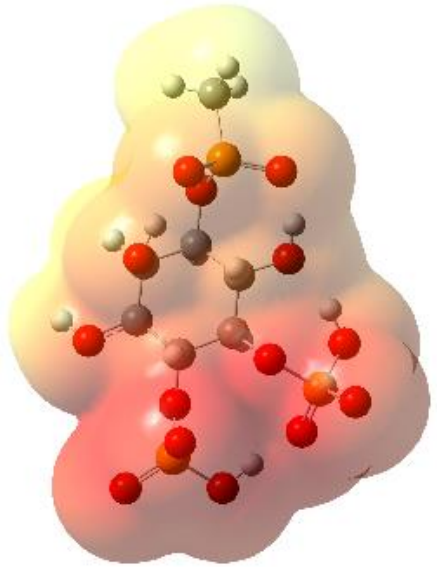 | 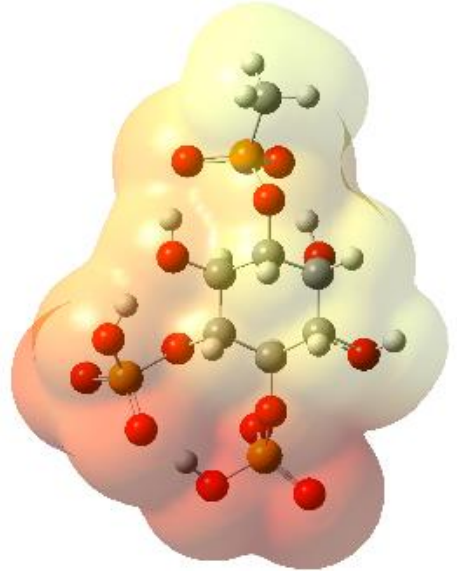 |
| 4a_PI34P | −2349.43641 | 24.661288 | 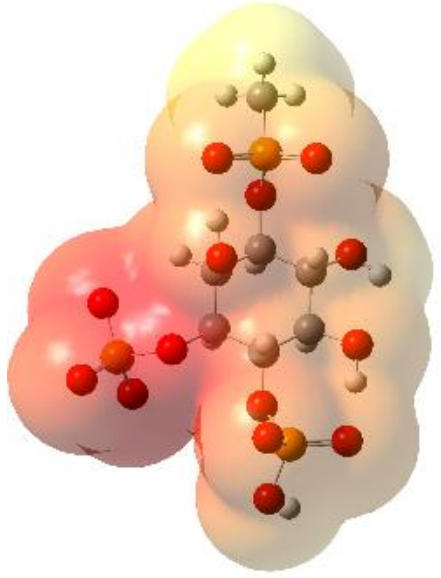 | 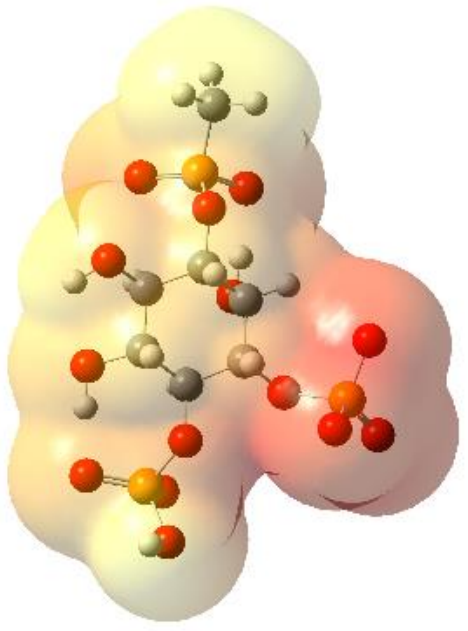 |
| 4b_PI34P | −2349.334163 | 23.013194 | 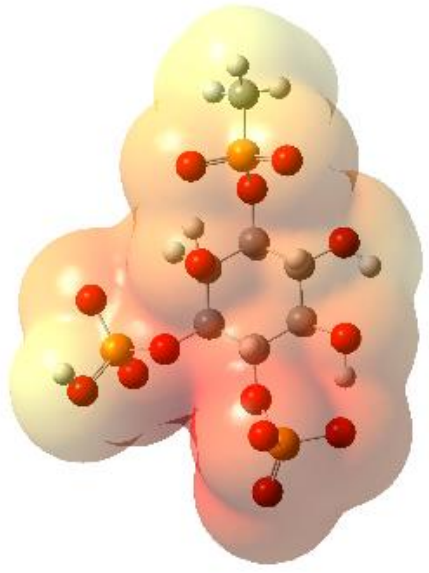 | 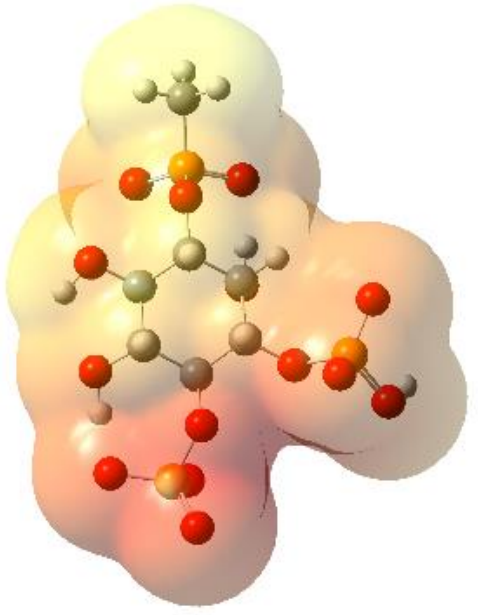 |
| 4a_PI35P | −2347.618004 | 27.058087 | 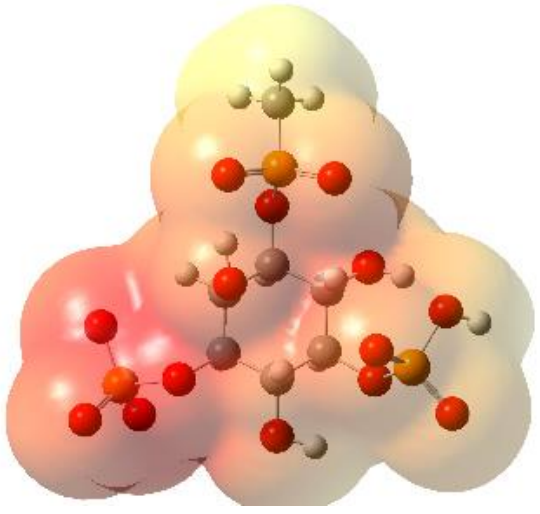 | 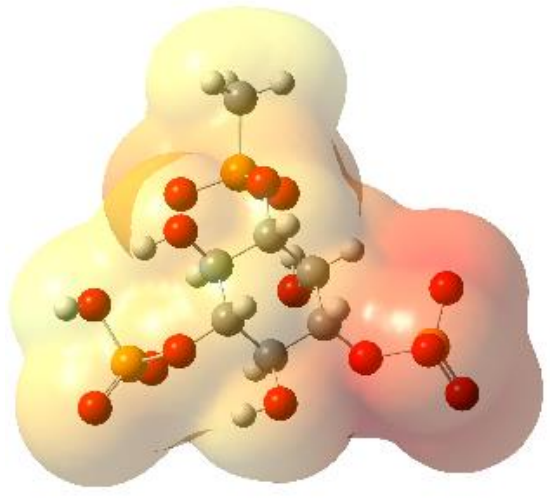 |
| 4b_PI35P | −2349.339481 | 15.841336 | 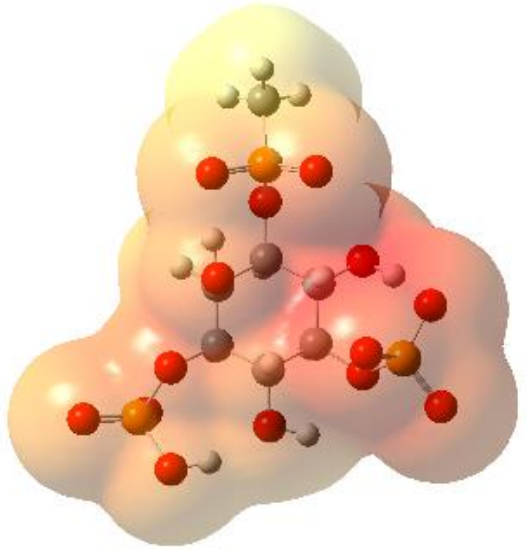 | 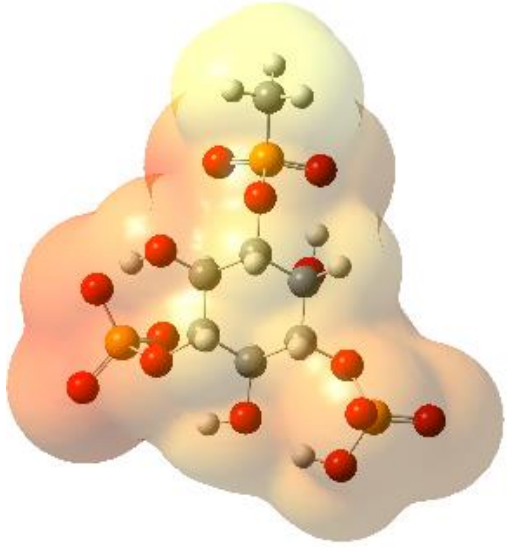 |
| 4a_PI45P | −2349.328898 | 41.322896 | 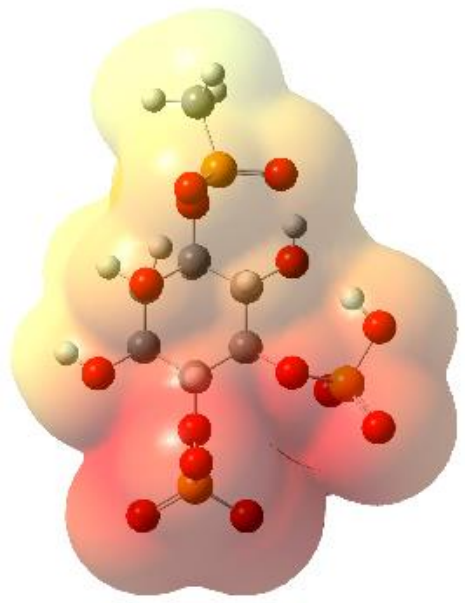 | 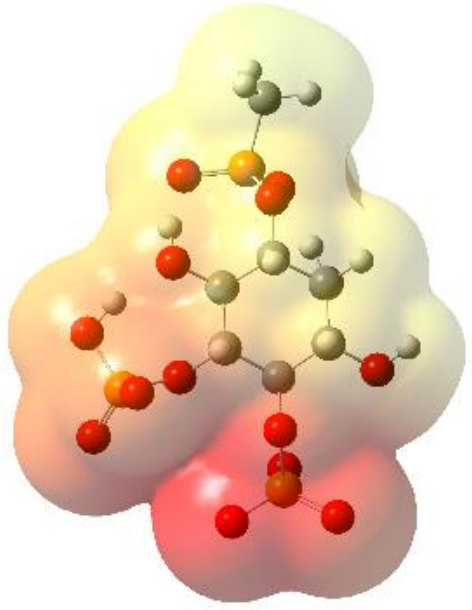 |
| 4b_PI45P | −2349.342221 | 33.688092 | 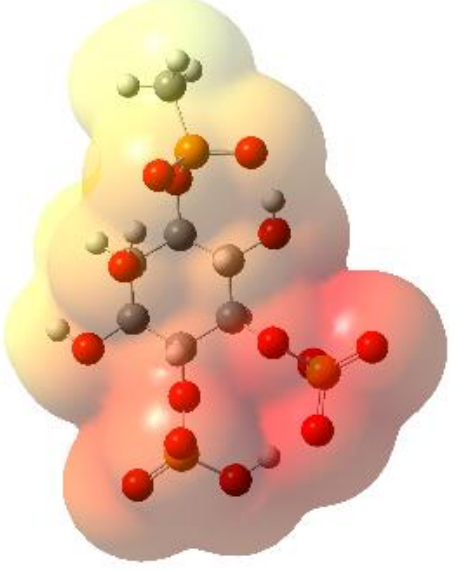 | 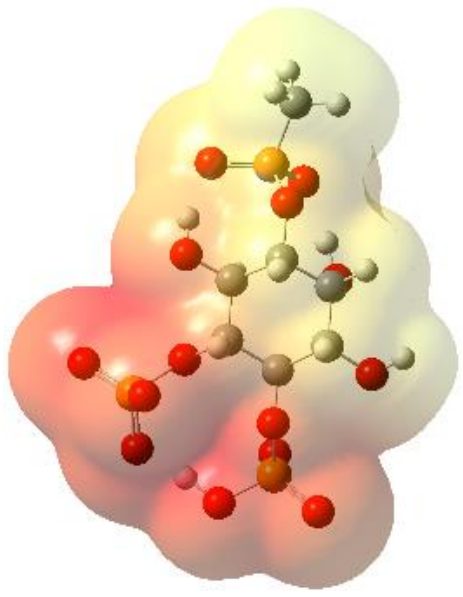 |
| 5_PI34P | −2348.832866 | 33.151083 | 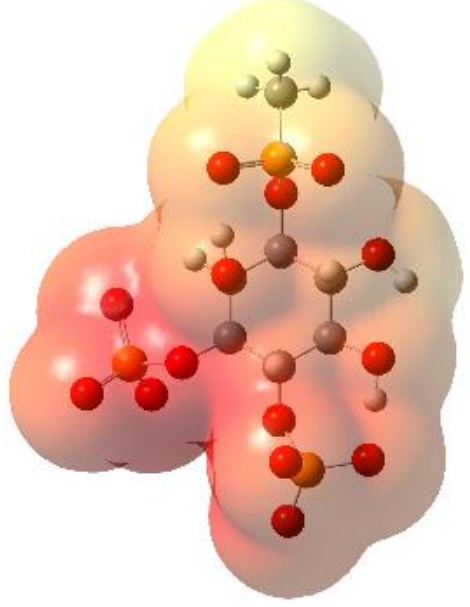 | 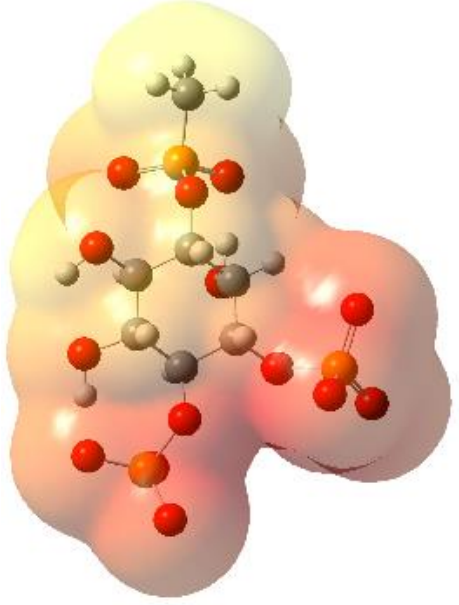 |
| 5_PI35P | −2348.830854 | 18.874273 | 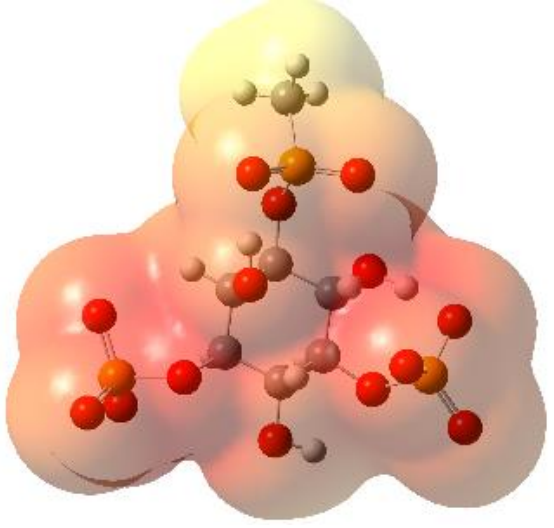 | 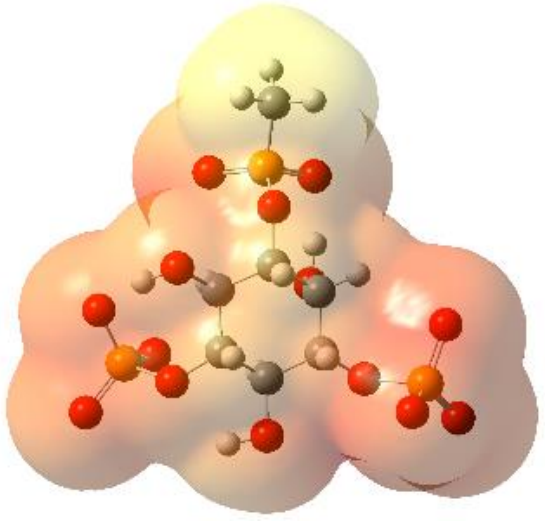 |
| 5_PI45P | −2348.808123 | 46.650622 | 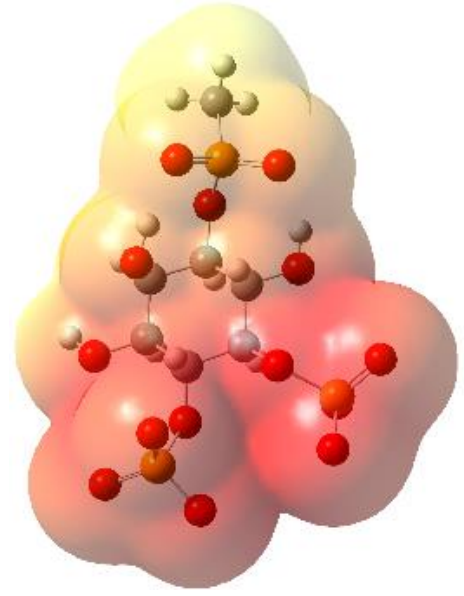 | 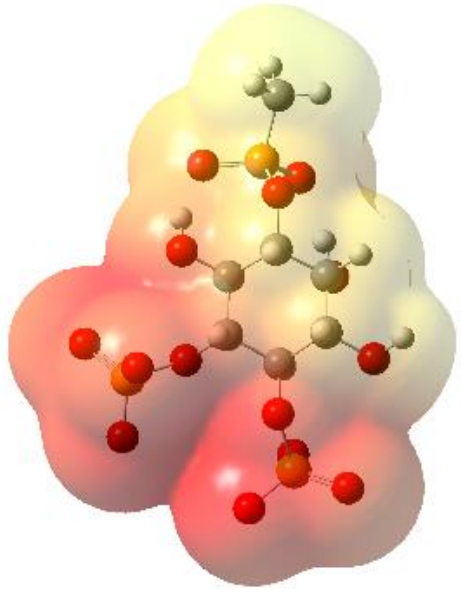 |
| 4_PI345P | −2916.40157 | 20.822726 | 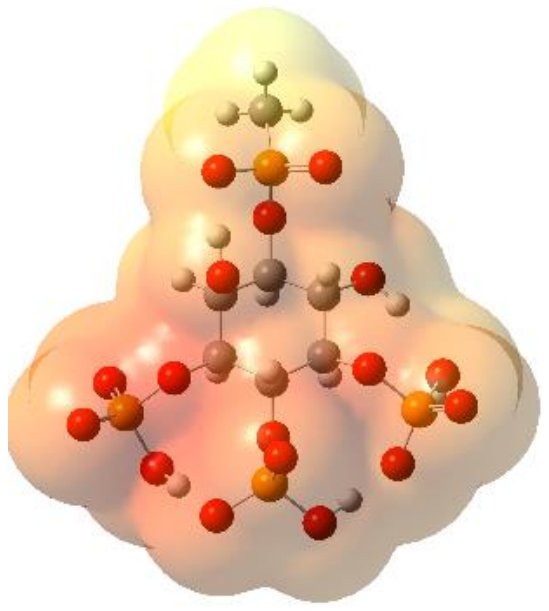 | 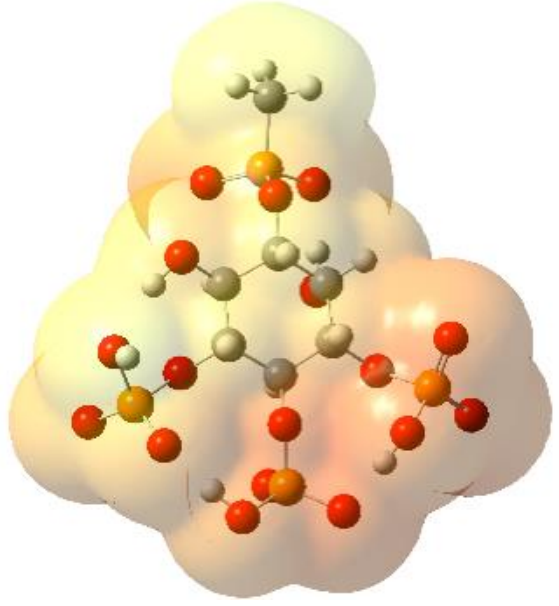 |
| 5a_PI345P | −2915.885588 | 31.906808 | 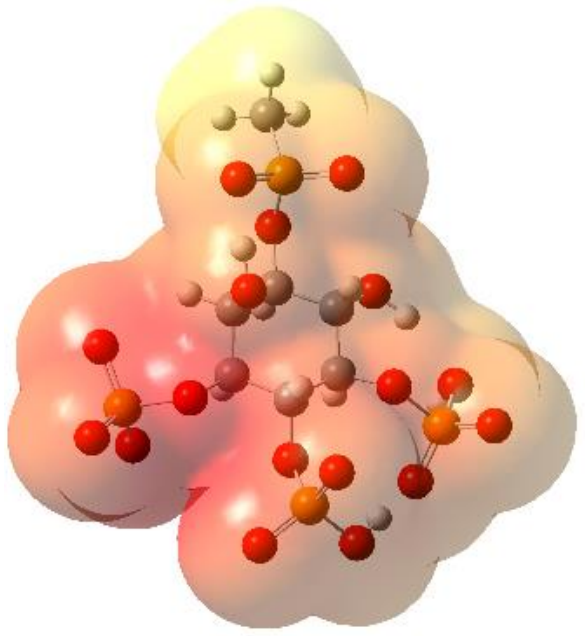 | 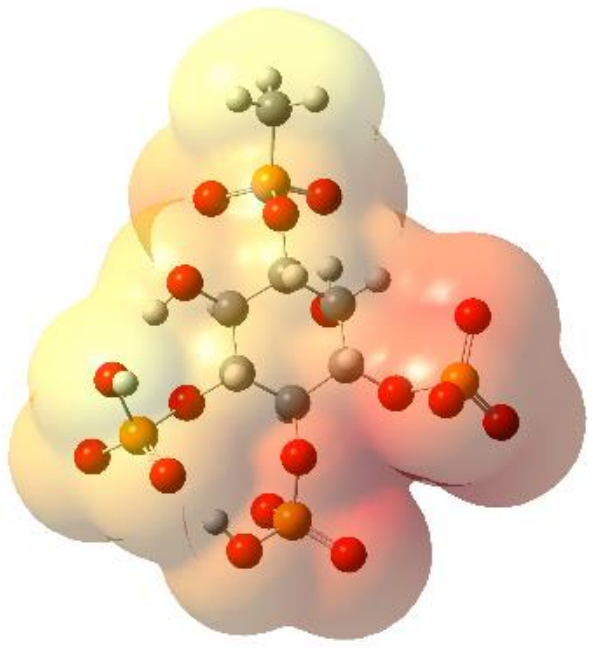 |
| 5b_PI345P | −2915.895338 | 31.5858 | 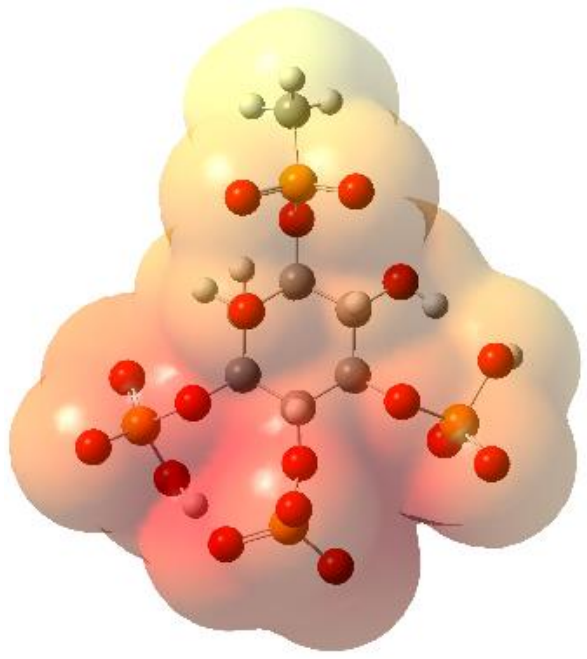 | 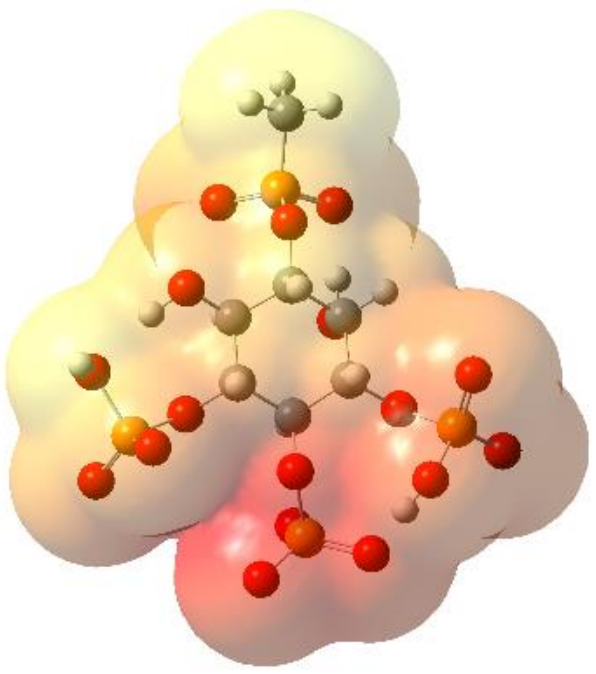 |
| 5c_PI345P | -2915.918291 | 19.277199 | 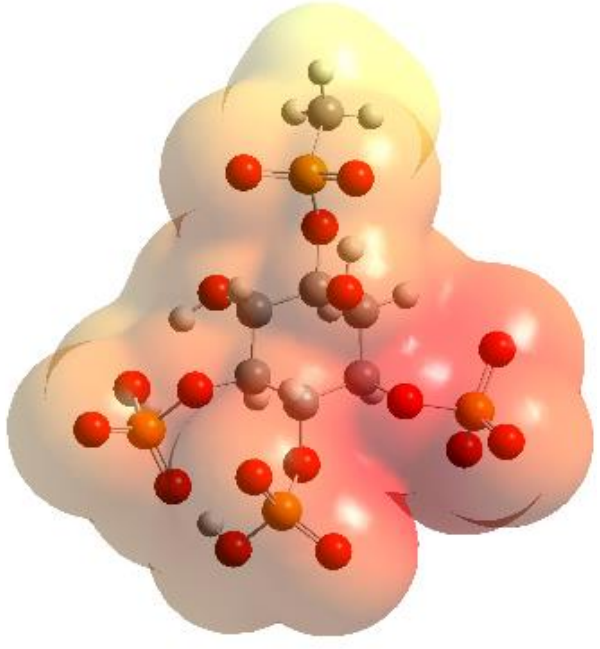 | 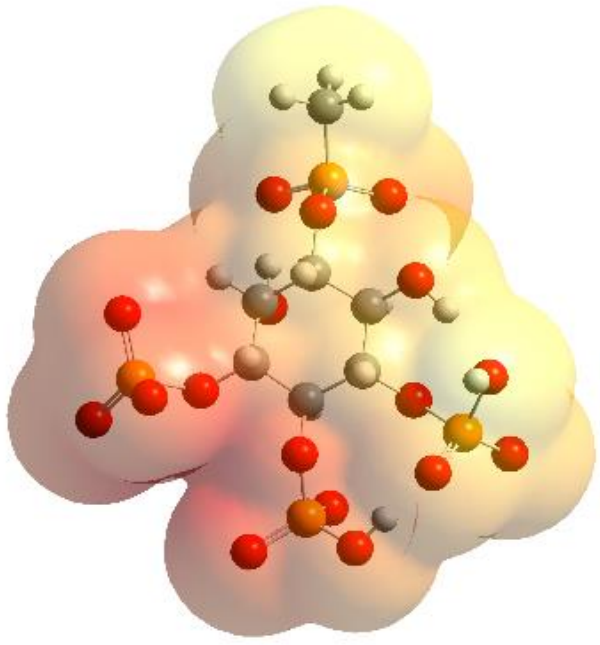 |
| 6a_PI345P | −2915.367158 | 42.506642 | 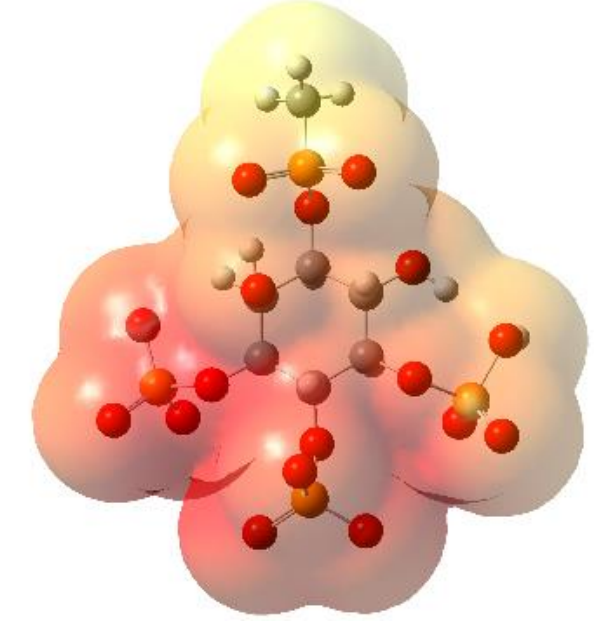 | 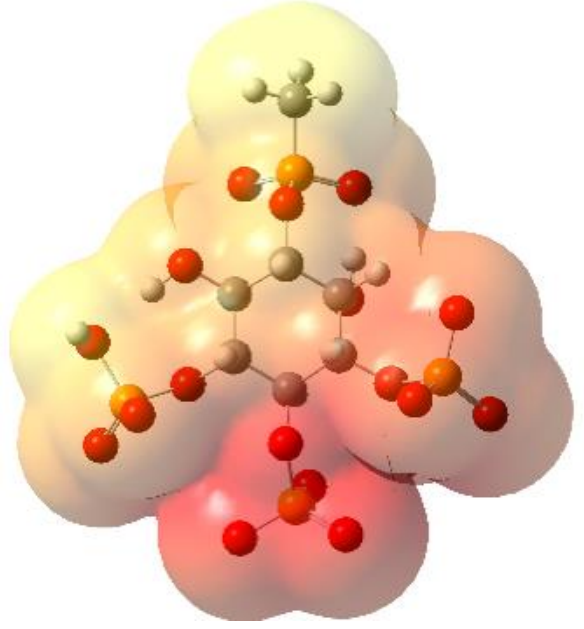 |
| 6b_PI345P | −2915.401568 | 25.249902 | 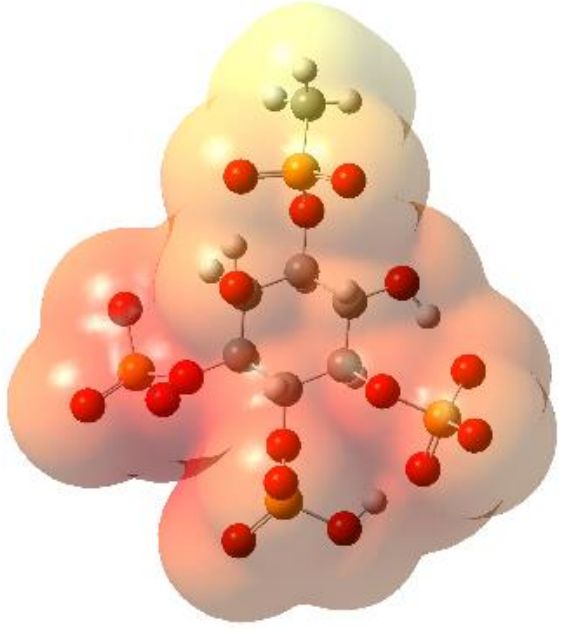 | 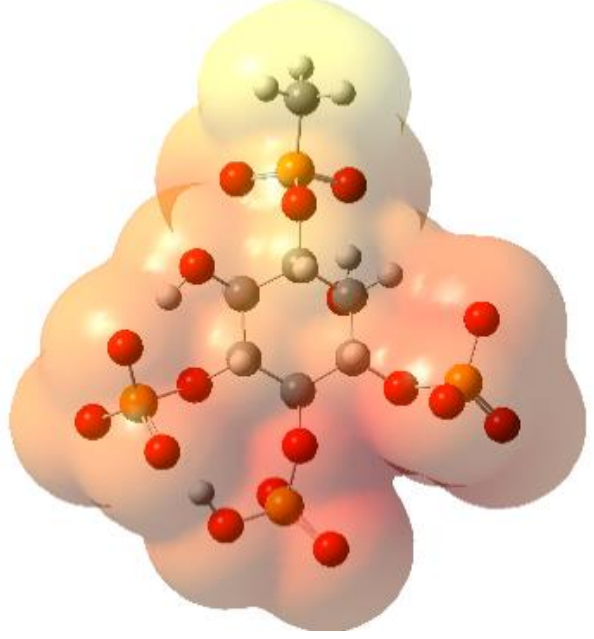 |
| 6c_PI345P | −2915.367159 | 42.497198 | 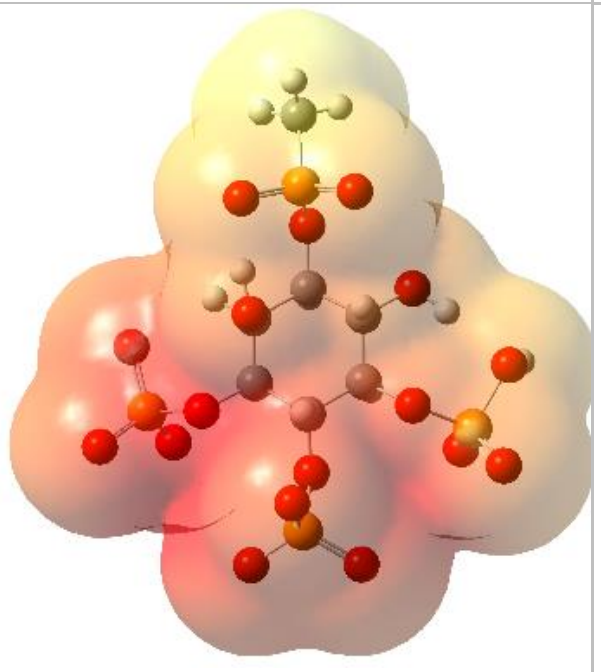 | 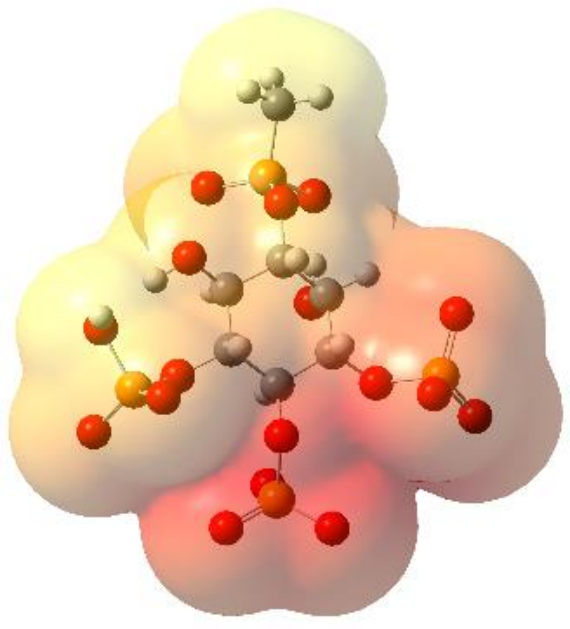 |
| 7_PI345P | −2914.873137 | 37.571612 | 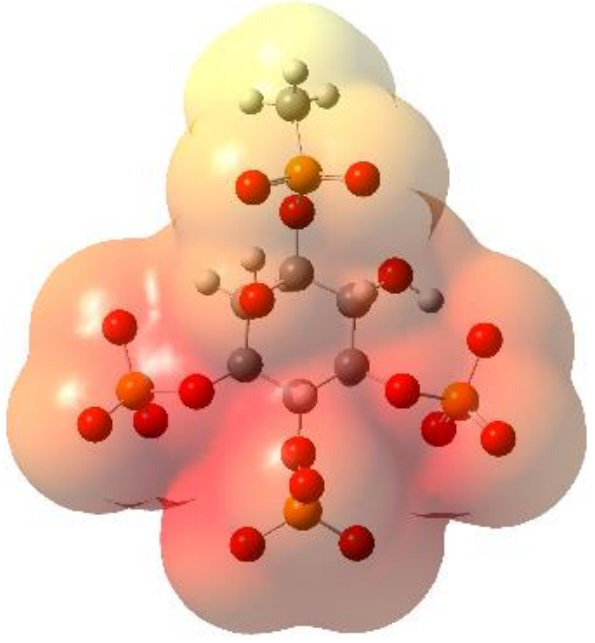 | 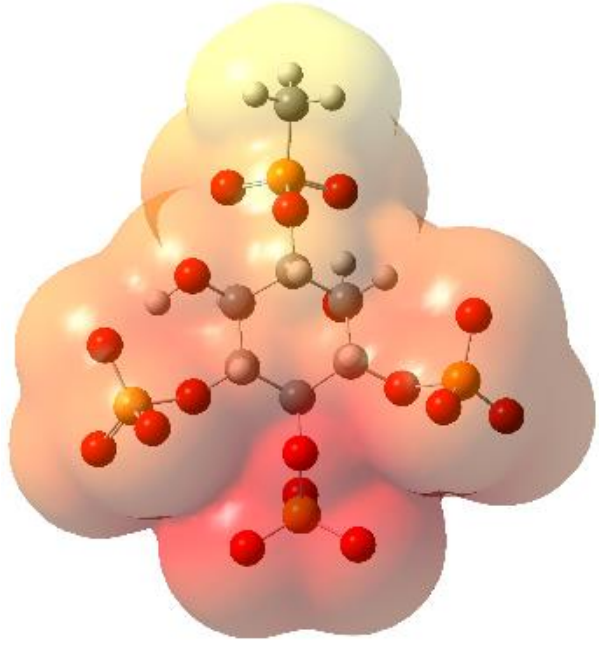 |

*The ESPs and RESPs for all 26 PIP headgroup structures, as well as their geometry optimisation files, will be available as a disk or via download from the Gould group public server. The file names corresponding to each structure are listed in* ***S.2*** *and* ***S.3*** *of Supporting Information:*

## S.2 ESP and RESP charges

| Structure names | ESP | | RESP | |
|---|---|---|---|---|
| | ESP charges | ESP surfaces | PDB files | Mol2 Files |
| 2_PI3P | 2_PI3Pesp.log | 2_PI3Pforamber.fchk<br>2_PI3Pforamber.cube<br>2_PI3Pforamber_ESP.cube | 2_PI3Pesp.pdb | 2_PI3Pesp.mol2 |
| 2_PI4P | 2_PI4Pesp.log | 2_PI4Pforamber.fchk<br>2_PI4Pforamber.cube<br>2_PI4Pforamber_ESP.cube | 2_PI4Pesp.pdb | 2_PI4Pesp.mol2 |
| 2_PI5P | 2_PI5Pesp.log | 2_PI5Pforamber.fchk<br>2_PI5Pforamber.cube<br>2_PI5Pforamber_ESP.cube | 2_PI5Pesp.pdb | 2_PI5Pesp.mol2 |
| 3_PI3P | 3_PI3Pesp.log | 3_PI3Pforamber.fchk<br>3_PI3Pforamber.cube<br>3_PI3Pforamber_ESP.cube | 3_PI3Pesp.pdb | 3_PI3Pesp.mol2 |
| 3_PI4P | 3_PI4Pesp.log | 3_PI4Pforamber.fchk<br>3_PI4Pforamber.cube<br>3_PI4Pforamber_ESP.cube | 3_PI4Pesp.pdb | 3_PI4Pesp.mol2 |
| 3_PI5P | 3_PI5Pesp.log | 3_PI5Pforamber.fchk<br>3_PI5Pforamber.cube<br>3_PI5Pforamber_ESP.cube | 3_PI5Pesp.pdb | 3_PI5Pesp.mol2 |
| 3_PI34P | 3_PI34Pesp.log | 3_PI34Pforamber.fchk<br>3_PI34Pforamber.cube<br>3_PI34Pforamber_ESP.cube | 3_PI34Pesp.pdb | 3_PI34Pesp.mol2 |
| 3_PI35P | 3_PI35Pesp.log | 3_PI35Pforamber.fchk<br>3_PI35Pforamber.cube<br>3_PI35Pforamber_ESP.cube | 3_PI35Pesp.pdb | 3_PI35Pesp.mol2 |
| 3_PI45P | 3_PI45Pesp.log | 3_PI45Pforamber.fchk<br>3_PI45Pforamber.cube<br>3_PI45Pforamber_ESP.cube | 3_PI45Pesp.pdb | 3_PI45Pesp.mol2 |
| 4a_PI34P | 4a_PI34Pesp.log | 4a_PI34Pforamber.fchk<br>4a_PI34Pforamber.cube<br>4a_PI34Pforamber_ESP.cube | 4a_PI34Pesp.pdb | 4a_PI34Pesp.mol2 |
| 4b_PI34P | 4b_PI34Pesp.log | 4b_PI34Pforamber.fchk<br>4b_PI34Pforamber.cube<br>4b_PI34Pforamber_ESP.cube | 4b_PI34Pesp.pdb | 4b_PI34Pesp.mol2 |
| 4a_PI35P | 4a_PI35Pesp.log | 4a_PI35Pforamber.fchk<br>4a_PI35Pforamber.cube<br>4a_PI35Pforamber_ESP.cube | 4a_PI35Pesp.pdb | 4a_PI35Pesp.mol2 |
| 4b_PI35P | 4b_PI35Pesp.log | 4b_PI35Pforamber.fchk<br>4b_PI35Pforamber.cube<br>4b_PI35Pforamber_ESP.cube | 4b_PI35Pesp.pdb | 4b_PI35Pesp.mol2 |
| 4a_PI45P | 4a_PI45Pesp.log | 4a_PI45Pforamber.fchk<br>4a_PI45Pforamber.cube<br>4a_PI45Pforamber_ESP.cube | 4a_PI45Pesp.pdb | 4a_PI45Pesp.mol2 |
| 4b_PI45P | 4b_PI45Pesp.log | 4b_PI45Pforamber.fchk<br>4b_PI45Pforamber.cube<br>4b_PI45Pforamber_ESP.cube | 4b_PI45Pesp.pdb | 4b_PI45Pesp.mol2 |
| 5_PI34P | 5_PI34Pesp.log | 5_PI34Pforamber.fchk<br>5_PI34Pforamber.cube<br>5_PI34Pforamber_ESP.cube | 5_PI34Pesp.pdb | 5_PI34Pesp.mol2 |
| 5_PI35P | 5_PI35Pesp.log | 5_PI35Pforamber.fchk<br>5_PI35Pforamber.cube<br>5_PI35Pforamber_ESP.cube | 5_PI35Pesp.pdb | 5_PI35Pesp.mol2 |
| 5_PI45P | 5_PI45Pesp.log | 5_PI45Pforamber.fchk<br>5_PI45Pforamber.cube<br>5_PI45Pforamber_ESP.cube | 5_PI45Pesp.pdb | 5_PI45Pesp.mol2 |
| 4_PI345P | 4_PI345Pesp.log | 4_PI345Pforamber.fchk<br>4_PI345Pforamber.cube<br>4_PI345Pforamber_ESP.cube | 4_PI345Pesp.pdb | 4_PI345Pesp.mol2 |
| 5a_PI345P | 5a_PI345Pesp.log | 5a_PI345Pforamber.fchk<br>5a_PI345Pforamber.cube<br>5a_PI345Pforamber_ESP.cube | 5a_PI345Pesp.pdb | 5a_PI345Pesp.mol2 |
| 5b_PI345P | 5b_PI345Pesp.log | 5b_PI345Pforamber.fchk<br>5b_PI345Pforamber.cube<br>5b_PI345Pforamber_ESP.cube | 5b_PI345Pesp.pdb | 5b_PI345Pesp.mol2 |
| 5c_PI345P | 5c_PI345Pesp.log | 5c_PI345Pforamber.fchk<br>5c_PI345Pforamber.cube<br>5c_PI345Pforamber_ESP.cube | 5c_PI345Pesp.pdb | 5c_PI345Pesp.mol2 |
| 6a_PI345P | 6a_PI345Pesp.log | 6a_PI345Pforamber.fchk<br>6a_PI345Pforamber.cube<br>6a_PI345Pforamber_ESP.cube | 6a_PI345Pesp.pdb | 6a_PI345Pesp.mol2 |
| 6b_PI345P | 6b_PI345Pesp.log | 6b_PI345Pforamber.fchk<br>6b_PI345Pforamber.cube<br>6b_PI345Pforamber_ESP.cube | 6b_PI345Pesp.pdb | 6b_PI345Pesp.mol2 |
| 6c_PI345P | 6c_PI345Pesp.log | 6c_PI345Pforamber.fchk<br>6c_PI345Pforamber.cube<br>6c_PI345Pforamber_ESP.cube | 6c_PI345Pesp.pdb | 6c_PI345Pesp.mol2 |
| 7_PI345P | 7_PI345Pesp.log | 7_PI345Pforamber.fchk<br>7_PI345Pforamber.cube<br>7_PI345Pforamber_ESP.cube | 7_PI345Pesp.pdb | 7_PI345Pesp.mol2 |

## S.3 Gaussian optimisation files

| Structure name | log files | chk files |
|---|---|---|
| 2_PI3P | 2_PI3Pforamber.log | 2_PI3Pforamber.chk |
| 2_PI4P | 2_PI4Pforamber.log | 2_PI4Pforamber.chk |
| 2_PI5P | 2_PI5Pforamber.log | 2_PI5Pforamber.chk |
| 3_PI3P | 3_PI3Pforamber.log | 3_PI3Pforamber.chk |
| 3_PI4P | 3_PI4Pforamber.log | 3_PI4Pforamber.chk |
| 3_PI5P | 3_PI5Pforamber.log | 3_PI5Pforamber.chk |
| 3_PI34P | 3_PI34Pforamber.log | 3_PI34Pforamber.chk |
| 3_PI35P | 3_PI35Pforamber.log | 3_PI35Pforamber.chk |
| 3_PI45P | 3_PI45Pforamber.log | 3_PI45Pforamber.chk |
| 4a_PI34P | 4a_PI34Pforamber.log | 4a_PI34Pforamber.chk |
| 4b_PI34P | 4b_PI34Pforamber.log | 4b_PI34Pforamber.chk |
| 4a_PI35P | 4a_PI35Pforamber.log | 4a_PI35Pforamber.chk |
| 4b_PI35P | 4b_PI35Pforamber.log | 4b_PI35Pforamber.chk |
| 4a_PI45P | 4a_PI45Pforamber.log | 4a_PI45Pforamber.chk |
| 4b_PI45P | 4b_PI45Pforamber.log | 4b_PI45Pforamber.chk |
| 5_PI34P | 5_PI34Pforamber.log | 5_PI34Pforamber.chk |
| 5_PI35P | 5_PI35Pforamber.log | 5_PI35Pforamber.chk |
| 5_PI45P | 5_PI45Pforamber.log | 5_PI45Pforamber.chk |
| 4_PI345P | 4_PI345Pforamber.log | 4_PI345Pforamber.chk |
| 5a_PI345P | 5a_PI345Pforamber.log | 5a_PI345Pforamber.chk |
| 5b_PI345P | 5b_PI345Pforamber.log | 5b_PI345Pforamber.chk |
| 5c_PI345P | 5c_PI345Pforamber.log | 5c_PI345Pforamber.chk |
| 6a_PI345P | 6a_PI345Pforamber.log | 6a_PI345Pforamber.chk |
| 6b_PI345P | 6b_PI345Pforamber.log | 6b_PI345Pforamber.chk |
| 6c_PI345P | 6c_PI345Pforamber.log | 6c_PI345Pforamber.chk |
| 7_PI345P | 7_PI345Pforamber.log | 7_PI345Pforamber.chk |